\documentclass[aps,prd,twocolumn,showpacs,superscriptaddress,nofootinbib,preprintnumbers]{revtex4-2}

\usepackage{style}
\usepackage{bbm}
\usepackage{subfiles}
\usepackage{amsfonts}
\usepackage[normalem]{ulem}
\usepackage{tikz}
\usetikzlibrary{arrows.meta, positioning}

\newcommand{\beq}{\begin{equation}}
\newcommand{\eeq}{\end{equation}}

\newcommand{\lcdm}{ $\Lambda$CDM }
\newcommand{\lya}{Lyman-$\alpha$ }

\newcommand{\kpeak}{k_{\rm{peak}}}
\newcommand{\kdamp}{k_{\rm{damp}}}

\def\lcdm{\Lambda \text{CDM}}
\newcommand*{\dif}{\mathop{}\!\mathrm{d}}

\begin{document}

\title{Strongest constraints on dark acoustic oscillations from the Lyman-alpha forest}

\author{Zhihan Yuan\,\orcidlink{0009-0007-4041-3861}}
\email{ly.yuan@mail.utoronto.ca}
\affiliation{Department of Physics, University of Toronto, 60 St. George Street, Toronto, ON, M5S 1A7, Canada}

\author{Caleb Gemmell\,\orcidlink{0000-0002-6505-8559}}
\email{cgemmell2@wisc.edu}
\affiliation{Department of Physics, University of Wisconsin, Madison, WI, 53706, USA}
\affiliation{Wisconsin IceCube Particle Astrophysics Center, University of Wisconsin, Madison, WI, 53706, USA}
\affiliation{Department of Physics, University of Toronto, 60 St. George Street, Toronto, ON, M5S 1A7, Canada}

\author{Keir K. Rogers}
\email{keir.rogers@kcl.ac.uk}
\affiliation{Department of Physics, King's College London, Strand, London, WC2R 2LS, UK}
\affiliation{Department of Physics, Imperial College London, Blackett Laboratory, Prince Consort Road, SW7 2AZ, UK}
\affiliation{Dunlap Institute for Astronomy and Astrophysics, University of Toronto, 50 St. George Street, Toronto, ON, M5S 3H4, Canada}

\author{Jared Barron\,\orcidlink{0000-0001-9043-6577}}
\email{jared.barron@stonybrook.edu}
\affiliation{C.N. Yang Institute for Theoretical Physics, Stony Brook University, Stony Brook, NY, 11794-3800, USA}

\author{Sandip Roy\,\orcidlink{0000-0002-7638-7454}}
\email{sandiproy@ucsd.edu}
\affiliation{Department of Astronomy and Astrophysics, University of California, San Diego, CA, 92093, USA}

\author{David Curtin\,\orcidlink{0000-0003-0263-6195}}
\email{dcurtin@physics.utoronto.ca}
\affiliation{Department of Physics, University of Toronto, 60 St. George Street, Toronto, ON, M5S 1A7, Canada}

\author{Norman Murray\,\orcidlink{0000-0002-8659-3729}}
\email{murray@cita.utoronto.ca}
\affiliation{Canadian Institute for Theoretical Astrophysics, University of Toronto, 60 St. George Street, Toronto, ON, M5S 3H8, Canada}
\affiliation{Department of Physics, University of Toronto, 60 St. George Street, Toronto, ON, M5S 1A7, Canada}

\begin{abstract}
We set the first constraints on a small-scale dark acoustic oscillation (DAO) in the linear matter power spectrum arising from dark sector interactions,  with a full forward model of the Ly-$\alpha$ forest. No more than 30\% of dark matter can form DAOs if they peak at wavenumbers $< 50\,h\,\mathrm{Mpc}^{-1}$ (95\% c.l.), probing scales $25 \times$ smaller than the cosmic microwave background (CMB). Given the complex covariance of DAO and nuisance parameters, we use a deep kernel learning emulator of hydrodynamical simulations to capture imprints of linear oscillations in the Ly-$\alpha$ forest.
\end{abstract}

\bigskip
\maketitle

%%%%%%%%%%%%%%%%%%%%%%%
%%%%%%%%%%%%%%%%%%%%%%%
%%%%%%%%%%%%%%%%%%%%%%%
%%%%%%%%%%%%%%%%%%%%%%%
%%%%%%%%%%%%%%%%%%%%%%%

\begin{figure}[!t]
    \centering
    \includegraphics[trim={0cm 0cm 0cm 0cm},clip,width=\columnwidth]{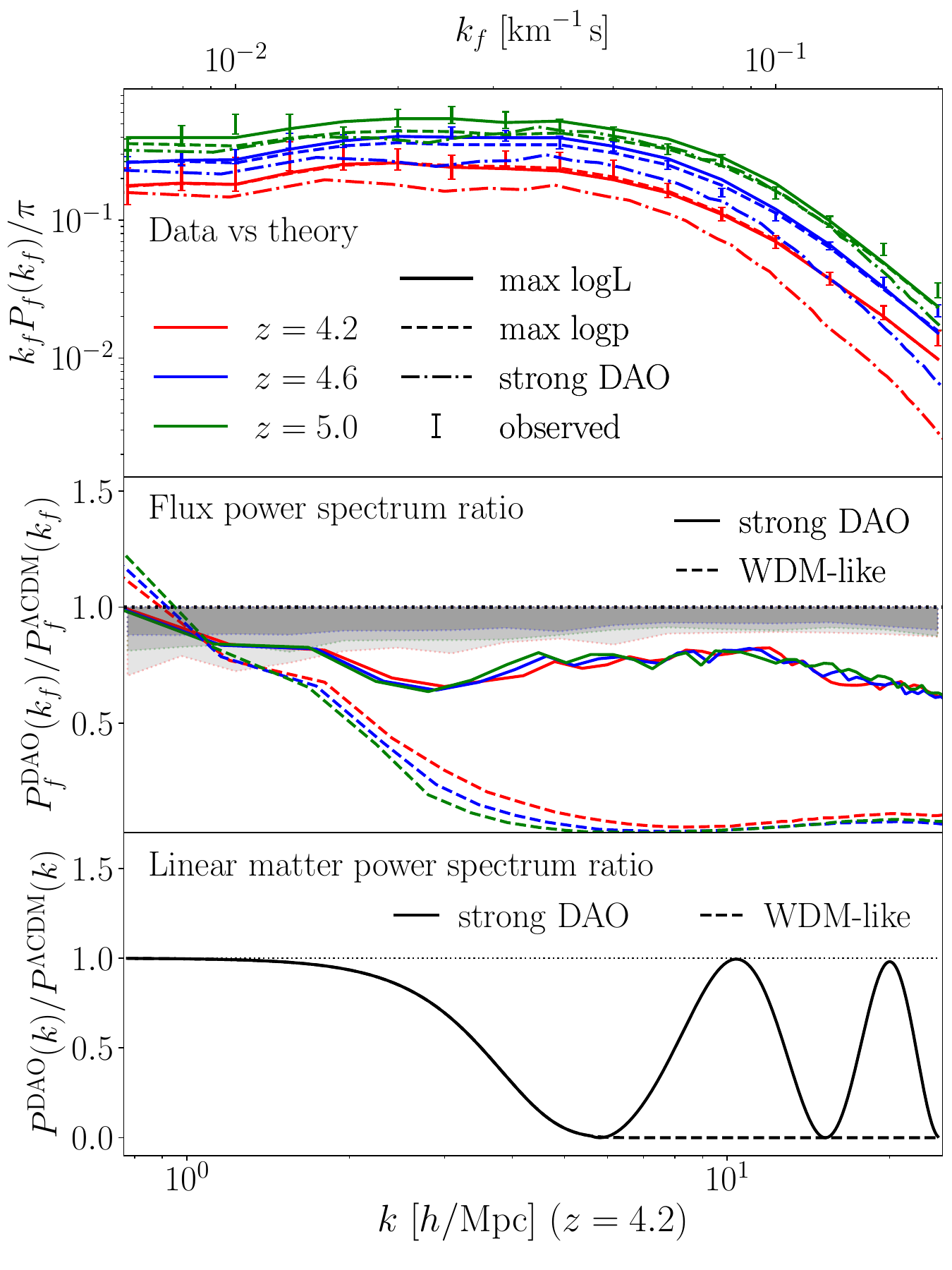}
    \caption{\textbf{Upper panel:} comparison of flux power spectra to data (points with errorbars). Red, blue, and green lines respectively show redshifts 4.2, 4.6, and 5.0. Solid lines are the maximum likelihood; dashed lines are the maximum posterior; and dot-dashed lines show a strong DAO model $(f,\kpeak,A)=(1,20\,h/\rm Mpc,1)$. \textbf{Middle panel:} ratio of flux power spectra to a $\Lambda$CDM model \((f=0)\). Solid lines are the same strong DAO model as above; dashed lines show a warm DM (WDM)-like model $(f,\kpeak,A)=(1,20\,h/\rm Mpc,0)$. The gray shaded regions indicate the data uncertainties (outlined by red, green, and blue according to their redshift). \textbf{Lower panel:} as middle panel, except for linear matter power spectra at \(z = 4.2\). } 
    \label{figure:excluded}
\end{figure}

\textbf{Introduction.} Discovering the fundamental nature of dark matter (DM) is one of the leading problems in the current era of physics. While the model of cold dark matter (CDM) is successful at explaining the large-scale structure of the Universe \cite{Planck:2018vyg}, it is largely unconstrained on smaller scales (\(\sim\) sub-Mpc). Additionally, the leading particle candidate for CDM, a weakly interacting massive particle, has only had consistent null results from collider and direct detection searches \cite{Yang:2016odq,LZ:2024zvo,XENON:2024xgd}. These results have motivated considerations for non-minimal models of DM which deviate from the small-scale structure predicted in $\Lambda$CDM. A distinctive feature 
of many such models is the presence of dark acoustic oscillations (DAOs) \cite{Kaplan:2009de,Cyr-Racine:2012tfp,Cyr-Racine:2013fsa}, which result in an overall suppression and oscillation feature in the Universe's matter power spectrum (see Fig.~\ref{figure:excluded}) 

DAOs can arise due to a coupling between DM and dark radiation in the dark sector, completely analogous to the process that causes baryon acoustic oscillations (BAOs) in $\Lambda$CDM. DAOs are additionally motivated as they 
can 
ameliorate discrepancies between observations and $\Lambda$CDM that have emerged in the last few years, the most significant being the Hubble, $S_{8}$ and CMB-BAO tensions \cite{Chacko:2016kgg,Bansal:2021dfh,Schoneberg:2021qvd, Buen-Abad:2022kgf,Garny:2025szk,Garny:2026ish}. 

One of the most theoretically motivated dark sectors giving rise to DAOs is atomic dark matter (aDM)~\cite{Kaplan:2009de}. An aDM subcomponent can cause deviations at larger~\cite{Cyr-Racine:2012tfp,Cyr-Racine:2013fsa,Bansal:2021dfh, Bansal:2021dfh,Barron:2025dys,Barron:2026nks}, smaller~\cite{Roy:2023zar,Gemmell:2023trd,Roy:2024bcu,MandacaruGuerra:2026zqk} and stellar~\cite{Curtin:2020tkm, Curtin:2019lhm, Hippert:2021fch, Armstrong:2023cis,Cabral:2026tjb,Shandera:2018xkn,Gurian:2021qhk,Gurian:2022nbx,Singh:2020wiq,Ryan:2022hku} scales, and is predicted by many models solving the Little Hierarchy Problem \cite{
Chacko:2005pe, Chacko:2016hvu, Craig:2016lyx,Barbieri:2016zxn,Arkani-Hamed:2016rle,Farina:2015uea,Alonso-Alvarez:2023bat}.  
However, since DAOs also arise in a wider class of dark sector models \cite{Chen_2002,Boehm_2005,Dvorkin_2014,Xu_2018,Gluscevic_2018,Fischer_2025}, we will focus on DAOs as a generic dark sector signature in this letter.

To this end, it is useful to implement an effective parameterization of DAOs in the linear matter power spectrum.
In the effective theory of structure (ETHOS) formalism \cite{Cyr_Racine_2016,Bohr:2020yoe}, the DAO transfer function (ratio of the linear matter power spectrum to the \(\Lambda\)CDM limit) is specified by the wavenumber and height of the first DAO peak, assuming a fully interacting dark sector. Further extending the range of theories covered by this parameterization, Ref.~\cite{Barron:2025dys} allowed the DM that interacts with the dark radiation to be a subset of the total DM abundance. We adopt
%implement
their model in this work.

Since DAOs imprint themselves on the matter power spectrum at small scales, the \lya forest, as a tracer of the small-scale (sub-Mpc), high-redshift (\(z \sim 5\)) linear matter power spectrum \cite{Cen:1994da,Croft:1997jf}, makes an excellent probe. In particular, high-resolution \lya forest spectra (which we use here) probe much smaller scales \cite{Bansal:2022qbi} than current cosmic microwave background (CMB) experiments like \textit{Planck} \citep{Planck:2018vyg}, the Atacama Cosmology Telescope \citep[ACT;][]{AtacamaCosmologyTelescope:2025blo,AtacamaCosmologyTelescope:2025nti,Lague:2026sbd} and the South Pole Telescope \citep{SPT-3G:2024atg}. Measurements of the high-redshift galaxy UV luminosity function (UVLF) have placed the strongest bounds on the DAO scale to date \cite{Barron:2025dys}, albeit given different priors than we use here. The \lya forest is 
%known to be 
a powerful probe of even smaller scales than the UV luminosity function, and so we use 
the \lya forest in this work to extend our sensitivity to DAO models.

The \lya forest is a spectral absorption signal arising from neutral hydrogen in the intergalactic medium (IGM), away from the highest-density regions of the cosmic web. This environment makes the \lya forest a powerful probe of DM behavior, as, at any given scale, it is less affected by non-linearities and galactic feedback processes than galaxy tracers. Previous studies have used the \lya forest to constrain alternative DM models such as warm dark matter (WDM) \cite{Viel:2013fqw,Irsic:2017ixq,Villasenor:2022aiy,Irsic:2023equ}, axion DM \citep{Rogers:2020ltq,Rogers:2020cup,Garcia-Gallego:2026phh}, interacting DM \cite{Rogers_2022}, decaying DM \citep{Liu:2020wqz,Capozzi:2023xie}, compact object DM \citep{Murgia:2019duy} and mixed DM models \cite{Kobayashi:2017jcf,Garcia-Gallego:2025kiw}. Further, the \lya forest appears to be uniquely sensitive to DAOs, as previously pointed out in Ref.~\cite{Bose:2018juc}. Existing work on the Lyman-$\alpha$ forest signature of dark matter models with DAOs has focused on a few benchmark models~\cite{Bose:2018juc}, the weak DAO limit~\cite{Krall:2017xcw,Garny:2018byk}, or ignored the oscillations, using a WDM-like model~\cite{Archidiacono:2019wdp}. Our work is the first to obtain constraints while fully accounting for the DAO feature as modeled through a suite of hydrodynamical simulations.

However, cosmological hydrodynamical simulations of the IGM are computationally expensive. A direct sampling of the parameter posterior distribution by such simulations is impractical. Accurate comparison of data to theory is enabled by machine learning (ML) emulator models and active learning methods \cite{2014ApJ...780..111H,Rogers:2018smb,Bird:2018efe,Rogers:2020cup,Rogers_2022}. 
The emulator model is trained on a limited set of simulations and is then used to interpolate the \lya forest flux power spectrum across the model parameter space. Active learning allows informed selection of new training simulation points after each iteration based on the knowledge gained from previous training sets. The emulator solution has been shown to accurately constrain cosmological, DM and IGM parameters using the \lya forest \cite{Rogers:2020ltq,Rogers_2022,Irsic:2023equ}.

In this work, we combine the probabilistic modeling used in previous works with deep learning methods. This hybrid approach, known as deep kernel learning (DKL), combines Gaussian processes (GPs) \citep{rasmussen_williams_2008} with neural networks.
We perform cosmological hydrodynamical simulations for different cosmological, IGM and DAO transfer function parameters. We train a DKL emulator that produces Lyman-$\alpha$ forest flux power spectra. We then use this emulator to place constraints in the DAO parameter space by performing a Bayesian inference with Lyman-$\alpha$ forest data \citep{Boera:2018vzq} collected from eleven quasar spectra from \textit{Keck}-HIRES \cite{1994SPIE.2198..362V} and four from VLT-UVES \cite{2000SPIE.4008..534D}.

\textbf{Methods.} We adopt the effective DAO transfer-function parameterization of Ref.~\cite{Barron:2025dys}. To minimize the computational cost of hydrodynamical simulations, we train an emulator on simulations selected via the Bayesian optimization active learning procedure of Ref.~\cite{Rogers:2018smb}. The resulting emulator predicts the Lyman-$\alpha$ forest flux power spectrum across the parameter space and is used to evaluate the likelihood, as part of Markov chain Monte Carlo sampling of the posterior.

\textit{DAO transfer function model~---}~We model the DAO feature in the linear matter power spectrum with the same phenomenological model as Ref.~\cite{Barron:2025dys}. The model has four free parameters: $f$, $A$, $\kpeak$, $\kdamp$. The suppression of the matter power spectrum is modeled by a mixed WDM and CDM-like transfer function \cite{Murgia:2017lwo}, with the depth of the suppression set by the fraction of the dark matter not in the CDM component $f$. The DAOs are modeled as a Gaussian-damped sinusoid whose first peak is at wavenumber $\kpeak$, with amplitude $A$ and the damping envelope set by $\kdamp$. $\kpeak$ also sets the wavelength of the oscillations. The bottom panel of Figure \ref{figure:excluded} shows examples of linear matter power spectra with strong DAOs ($A$=1) and a WDM-like model with no DAOs ($A$=0).

Since the simulations are expensive, it is preferable to minimize the dimensionality of the parameter space. We fix $\kdamp = 10 \kpeak$, a conservative approximation for setting constraints, which we justify in Appendix \ref{app:transfer_function}, reducing our DAO model to three parameters $\{f,A,\kpeak\}$.

For the DAO parameters, we assume uniform priors: $k_{\rm peak}\sim\mathcal{U}(0.2\: h\,\mathrm{Mpc}^{-1},50\, h\,\rm Mpc^{-1})$, $A\sim\mathcal{U}(0,1.5)$, and $f\sim\mathcal{U}(0,1)$. The range of $f$ spans the full physically allowed interval by construction, while the range of $A$ encompasses values expected in viable atomic dark matter models. The upper bound on $k_{\rm peak}$ is chosen because the Lyman-$\alpha$ forest becomes increasingly insensitive to DAO features at smaller scales, preventing meaningful constraints beyond $k_{\rm peak}\simeq 50\: h\,\rm Mpc^{-1}$.

\textit{Cosmological and intergalactic medium parameters~---}~
The matter power spectrum inferred from the Lyman-$\alpha$ forest is also sensitive to the cosmological model and the thermalization and ionization of the IGM, primarily due to the uncertain nature of reionization. To ensure our constraints are robust, we thus marginalize over cosmological and IGM parameters that capture these uncertainties.

For the cosmological parameters, we vary the spectral index $n_s$ and amplitude $A_s$ of the primordial power spectrum, assuming Gaussian priors with means and standard deviations $(0.9649,\,0.004)$ and $(2.101\times10^{-9},\,2.962\times10^{-11})$, respectively, derived from \textit{Planck} CMB results \cite{Planck:2018vyg}. The Lyman-\(\alpha\) forest flux power spectrum is not sensitive to other standard cosmological parameters \citep{Pedersen:2019ieb,Rogers:2023upm}. The transfer function parameters only capture the effects of the DAOs due to the coupled plasmas. We thus also vary the effective number of additional relativistic species $\Delta N_\mathrm{eff}$ 
to ensure the effects of free-streaming radiation on the matter power spectrum are considered. We assume a uniform prior on the dark sector temperature ratio $\xi \sim \mathcal{U}(0, 0.5)$, which relates to $\Delta N_\mathrm{eff} = \frac{8}{7}(\frac{11}{4})^{4/3}\xi^4 
\approx 4.4 \ \xi^4$.

To marginalize over uncertainty in the properties of the IGM, we vary the simulation input parameters $\{H_A, H_S, z_{\rm{rei}}, T_{\rm{rei}}\}$, which modify baseline ultraviolet background (UVB) rates, taken from Ref.~\cite{Faucher-Giguere:2019kbp}. $H_A$ is a multiplicative factor, while $H_S$ applies an overdensity \(\Delta\)-dependent rescaling, such that the new photoheating rates implemented are $\epsilon_{i} = H_A \times \epsilon_{0,i} \times \Delta^{H_S}$, where $\epsilon_{0,i}$ are the default photoheating rates, with $i \in $ [HI, HeI, HeII]. We additionally modify the default photoionization rates by varying the reionization redshift $z_{\rm{rei}}$ and total heat injection $T_{\rm{rei}}$ as implemented in Ref.~\cite{Onorbe:2016hjn}. Thus, for the IGM, we vary $H_A \in$ [0.05, 2.5], $H_S \in$ [-1, 0.7], $T_{\rm{rei}} \in$ [$1.5\times10^4$, $4\times10^4$] K, $z_{\rm{rei}} \in$ [6, 7.8], with the maximum $z_{\rm{rei}}$ value set by the default value in the baseline photoionization rates.

While the above parameters are varied as simulation inputs, we use the following output parameters to characterize the IGM in each simulation $\{T_0(z = z_i),\tilde\gamma(z = z_i),u_0(z = z_i),\tau_0(z = z_i)\}$, for each redshift that we consider \(z_i = [4.2, 4.6, 5.0]\). The vast majority of neutral gas in the IGM follows a temperature \(T\)-density relation: $T(z)=T_0(z)\Delta^{\tilde\gamma(z)-1}$. We thus describe the IGM thermal state using $T_0(z)$, $\tilde{\gamma}(z)$, and also $u_0 (z)$, the cumulative thermal energy
injected into the IGM per unit mass until redshift $z$, in units of
${\rm eV}\,m_p^{-1}$. We further vary a redshift-dependent normalization
$\tau_0(z)$ of the effective optical depth,
$\tau_{\rm eff}(z)=\tau_0(z)\tau_{\rm eff}^{\rm fid}(z)$, with
$\tau_{\rm eff}\equiv -\ln\langle F\rangle$ defined by the mean transmitted
flux fraction \(F\). For \(\tau_\mathrm{eff}^\mathrm{fid}\), we use the model of Ref.~\cite{Boera:2018vzq}. Following previous work \citep{Irsic:2017ixq,Kobayashi:2017jcf,Murgia:2018now,Rogers:2020ltq,Rogers_2022} and to disfavor unphysically cold IGMs with an inverted temperature-density relation, Gaussian priors are adopted for the IGM parameters:
$\tau_0(z)\sim\mathcal{N}(1.0,0.05^2)$,
$\gamma(z)\sim\mathcal{N}(1.2,0.2^2)$, and
$T_0(z)\sim\mathcal{N}((8022,7651,8673)\,\mathrm{K},(3000\,\mathrm{K})^2)$
at $z=(5.0,4.6,4.2)$, respectively. We additionally impose a uniform prior within the convex hull of the IGM simulations to prevent unphysical combinations of \(T_0\) and \(u_0\) and prevent unphysical jumps in \(T_0\) greater than 5000 K and in \(u_0\) greater than \(10\,\mathrm{eV}\,m_\mathrm{p}^{-1}\) from each redshift bin to the next.

\textit{Cosmological hydrodynamical simulations~---~} 
Our observable is the Lyman-$\alpha$ forest 1D flux power spectrum at $z=[4.2, 4.6, 5.0]$, i.e., the two-point line-of-sight correlation in the transmitted flux contrast in Fourier space. In order to model this quantity, we run cosmological hydrodynamical simulations of the IGM using the publicly-available code \texttt{GIZMO} \cite{Hopkins:2014qka}. We first produce \(\Lambda\)CDM-like linear matter power spectra using the Boltzmann code \texttt{CLASS-aDM} \cite{Bansal:2022qbi}, which are then modified by the DAO transfer function as described in Appendix \ref{app:transfer_function}. The initial conditions of the simulations are generated using this modified matter power spectrum
and \texttt{MUSIC} \cite{10.1111/j.1365-2966.2011.18820.x} at $z=99$. We generate separate initial conditions for the dark matter and baryonic components \citep{Fernandez:2020jgf}. We then evolve $512^3$ particles each of dark matter and gas in a periodic $(10\, h^{-1}$ Mpc)$^3$ box from $z = 99$ to $z = 4.2$, saving snapshots of particle data at $z = [4.2, 4.6, 5.0]$.  As we explain in Appendix~\ref{app:cooling_approx}, we do not need to include the hydrodynamical effects of non-minimal DM in our simulations.

To reduce significantly the computational expense of each simulation, but with negligible effect on the flux power spectrum, we implement a simplified star formation criterion (\texttt{QuickLymanAlpha}), following Ref.~\cite{Viel:2004bf}: gas particles at overdensities $>$ 1000 and with temperatures $<$ $10^5$ K are converted to collisionless star particles. We verify this approach in Appendix \ref{app:fast_flag}, where we also confirm insensitivity to the choice of feedback model.

From each particle snapshot, we generate 32000 mock quasar spectra along one axis (with spectral pixel widths $\Delta v = 1\,\mathrm{km}\,\mathrm{s}^{-1}$) of the \lya absorption line and then calculate the 1D (line-of-sight) flux power spectrum using \texttt{fake\_spectra} \cite{2017ascl.soft10012B}. These 1D flux power spectra are generated at $z=[4.2, 4.6, 5.0]$ for each parameter point and used as training data for the emulator (\textit{below}). A total number of 413 simulations were used for training. For each simulation, we vary the mean flux rescaling $\tau_0$ and produce ten flux power spectra, which increases the size of the training dataset to 4130.
We perform tests of numerical convergence with respect to particle number and box size
in Appendix \ref{app: convergence}.
We correct for the box size effect by rescaling the flux power spectra. The rescaling ratios used are 0.979 for $z=5$, 0.959 for $z=4.6$, and 0.940 for $z=4.2$.

\textit{Deep kernel learning emulator and Bayesian optimization~---~}We emulate the flux power spectrum as a function of the eighteen model parameters $\bm{\theta}=\{k_{\rm{peak}},A,f,n_s,A_s,\xi,T_0(z=z_i),\tilde\gamma(z=z_i),u_0(z=z_i),\tau_0(z=z_i)\}$, for \(z = [4.2,4.6,5.0]\). 
The emulator is trained on our Lyman-$\alpha$ forest simulations (\textit{above}). Previous work has used Gaussian process emulators for Lyman-$\alpha$ forest flux power spectra \cite{Rogers:2018smb,Bird:2018efe,Rogers:2020ltq,Rogers:2020cup,Pedersen:2020kaw,Rogers_2022,Bird:2023evb}. A GP emulator interpolates simulation outputs by modeling their covariance through a kernel function \citep{rasmussen_williams_2008}. Standard kernels assume stationarity, i.e., that correlations depend only on the separation between points in parameter space. However, the response of the Lyman-$\alpha$ forest flux power spectrum to DAO parameters is highly nonstationary.

When $f=0$, there are no DAO effects, regardless of $\kpeak$ and $A$. As $f$ increases, $\kpeak$ and $A$ begin to influence the flux power spectrum, although their effects are still highly dependent on $f$. We thus use deep kernel learning (DKL) emulation \cite{wilson16}, as implemented in \texttt{GPyTorch}. The DKL emulator combines a neural network (NN) with a GP, using the NN as a feature extractor to map the original parameter space into a latent representation where a stationary kernel can accurately model the covariance structure.

The NN feature extractor is physically informed, guiding the learned latent representation toward physically relevant parameter combinations and improving emulator performance. The feature extractor and GP are trained jointly until convergence, defined by the change in the NN hyperparameters falling below a prescribed threshold. Then, we freeze the NN hyperparameters and fine-tune the GP kernel with the evidence lower bound (ELBO) as the loss function~\cite{leibfried2022tutorialsparsegaussianprocesses}. The DKL emulator allows us to represent the non-stationarity and degeneracies in our parameters while retaining the predictive variance from GP models.
Because a full GP scales poorly with the size of the training set, retraining the GP during optimization of the feature extractor would be computationally prohibitive. We therefore employ a variational GP, which represents the latent space covariance through a small set of learnable inducing points. These inducing points act as a compressed representation of the training data and are optimized jointly with the neural network and GP.
For a detailed explanation, see
Appendix \ref{app: emulator}.

The initial emulator was constructed using 68 parameter points arranged in a Latin hypercube. From this initial set, we used Bayesian optimization (as previously used by Refs.~\cite{Rogers:2018smb,Rogers:2020ltq,Rogers_2022}) to iteratively add simulations to the training set. This approach uses observed data and emulator uncertainty quantification from the GP to decide the optimal construction of the training set. We ran 413 simulations in total, and stopped adding training simulations once the estimated posterior distribution  converged with respect to the training set (see Appendix \ref{app: emulator} for details).

\textit{Posterior distribution sampling~---~}We compare the emulator output to the flux power spectrum measured in Ref.~\cite{Boera:2018vzq}, derived from eleven quasar spectra from \textit{Keck}-HIRES \cite{1994SPIE.2198..362V} and four from VLT-UVES \cite{2000SPIE.4008..534D}, which contains the smallest scales measured to-date in the Lyman-$\alpha$ forest (velocity wavenumbers \(k_f < 0.2\,\mathrm{s}\,\mathrm{km}^{-1}\); see Fig.~\ref{figure:excluded}). We assume a Gaussian likelihood function which includes emulator uncertainty. We use Markov chain Monte Carlo sampling to estimate the posterior distribution using the \texttt{emcee} sampler \citep{ForemanMackey:2013}, declaring convergence once each chain is fifty times the auto-correlation length.

\begin{figure}
    \centering
    \includegraphics[trim={0cm 0cm 0cm 0cm},clip,width=\columnwidth]{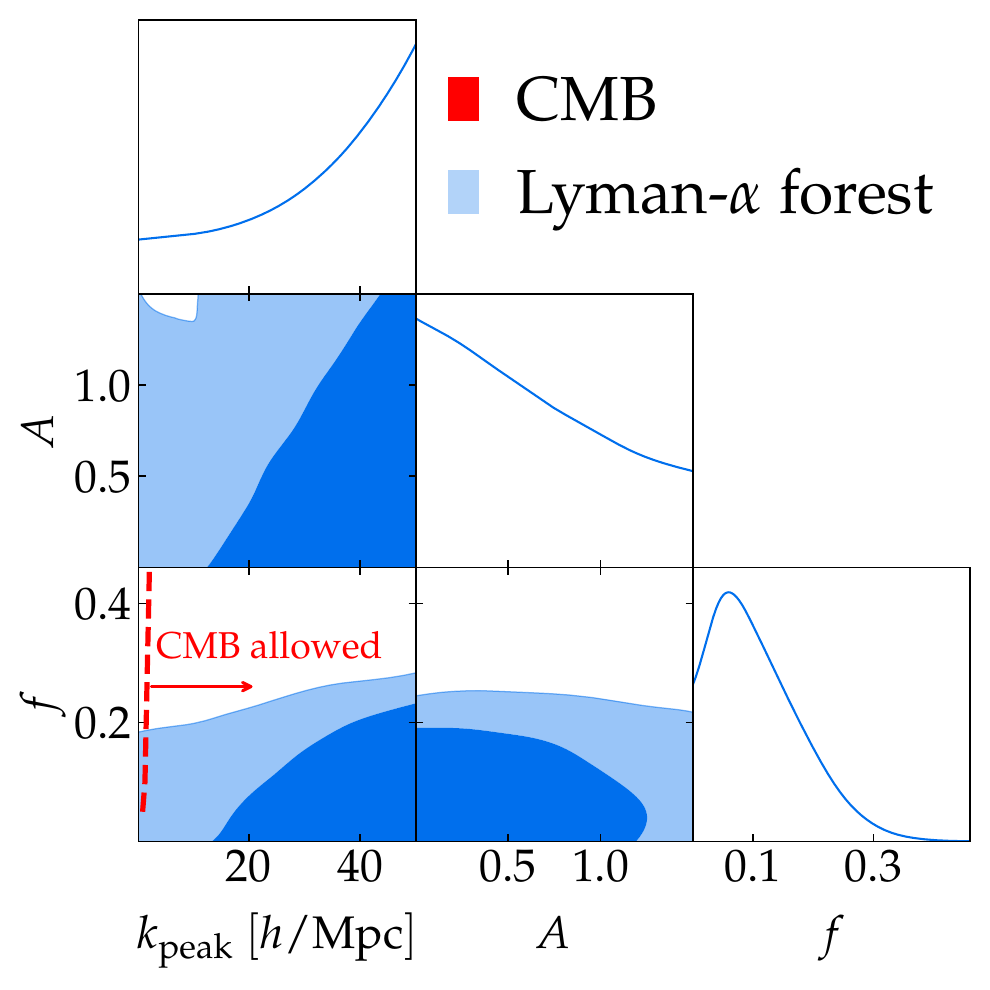}
        \caption{Blue contours show the 2D marginalized posterior (i.e., allowed region) of DAO transfer function parameters given Lyman-\(\alpha\) forest data. The darker and lighter shaded areas indicate respectively the 68\% and 95\% credible regions. The 1D $95\%$ credible bounds on the DAO parameters are $k_\mathrm{peak} > 5.10~h/\mathrm{Mpc}$, $A < 1.38$, $f < 0.243$. The red dashed line in the $\kpeak$-$f$ panel indicates CMB bounds~\cite{Barron:2026nks}.}
    \label{figure:constraints}
\end{figure}

\textbf{Results and discussion.} 
The lower panel of Fig.~\ref{figure:excluded} shows the ratio of the linear matter power spectra of models with strong DAOs ($f=1$, $\kpeak=20\:h$/Mpc, $A=1$) and WDM-like suppression ($f=1$, $\kpeak=20\:h$/Mpc, $A=0$) to the corresponding $\Lambda$CDM model ($f=0$), with all other cosmological and IGM parameters fixed. The matter power spectrum ratio equals the square of the transfer function \(T(k)\) (see Appendix \ref{app:transfer_function}), so the first (negative) DAO trough in \(T(k)\) becomes the first DAO peak in the power spectrum, while the second DAO peak is at $\kpeak$. The middle panel of Fig.~\ref{figure:excluded} shows the imprint of these two models on the flux power spectrum at the three redshifts we consider. As previously seen, a small-scale WDM-like linear matter power spectrum suppression causes a strong small-scale suppression in the flux power spectrum. We demonstrate that the addition of a strong DAO, however, significantly changes the flux power. In particular, the oscillatory feature largely washes out, but it does reinstate flux power at all scales below the initial cut-off. We thus anticipate that the Lyman-\(\alpha\) forest will be sensitive to DAO features beyond simple suppressions and enhancements.

The upper panel in Fig.~\ref{figure:excluded} compares emulated flux power spectra to observations. The dot-dashed lines show the same strong DAO model as above, illustrating that such features at the scales considered are strongly disfavored. The dashed lines show the emulated flux power spectra at the maximum posterior parameter point, while the solid lines show those for the best-fit point. These points are not identical owing to prior volume effects. The fit is in general good, apart from the final wavenumber bin, where the flux power is underestimated. This result is consistent with previous analyses \citep[e.g.,][]{Irsic:2023equ}, who attributed this feature to either the noise model or a signature of power enhancement beyond \(\Lambda\)CDM. We investigate this feature further in Appendix \ref{app:systematic_checks}, finding that it does not affect our conclusions since the dominant DAO sensitivity derives from the 
largest affected scales.
% logic: emphasizing that the sensitivity comes from the largest scales in the data naively contradicts our claim that we're using LyA to probe DAO at 'smaller scales than before'

Figure \ref{figure:constraints} shows the 2D posterior distribution of the DAO parameters $\kpeak$, $A$ and $f$ given observed \lya forest flux power spectra after marginalizing over all cosmological and IGM parameters, compared to the parameter space previously allowed by the CMB. The $\kpeak$-$A$ panel demonstrates that Lyman-$\alpha$ forest data are indeed sensitive to the presence of DAOs. The data disfavor strong DAOs (larger \(A\)) when $\kpeak$ is smaller. The \(\kpeak\) - \(f\) panel demonstrates that the Lyman-$\alpha$ forest limits the abundance of DM forming DAOs to be \(f \lesssim 0.2\) when \(\kpeak \sim 0.2\,h\,\mathrm{Mpc}^{-1}\), but that this limit relaxes to \(f \lesssim 0.3\) when \(\kpeak \sim 50\,h\,\mathrm{Mpc}^{-1}\). Otherwise, we find that the data are consistent with the \(\Lambda\)CDM limit (\(f = 0\)). The uniform prior on \(\kpeak\) means that we do not sample the other \(\Lambda\)CDM limit when \(\kpeak \to \infty\); however, we find that these data lose sensitivity to DAOs (i.e., the parameter \(A\)) already at the maximum \(\kpeak\) that we consider. We discuss parameter degeneracies between DAO and cosmological and IGM parameters in Appendix \ref{app:degeneracies}. The 1D marginalized 95\% credible intervals are given in Table \ref{tab:credible_intervals} in Appendix \ref{app:degeneracies}.

Studies of the impact of strong DAOs on the CMB~\cite{Cyr-Racine:2012tfp,Bansal:2022qbi,Barron:2026nks} have worked directly in the parameter space of atomic dark matter, yielding constraints that are most clearly expressed in terms of the fraction of dark matter that is atomic, the dark photon temperature, and the dark sound horizon. Translated into the parameters of our model, the most recent constraints with \textit{Planck} and ACT DR6 data~\cite{AtacamaCosmologyTelescope:2025blo} reach down to the $f\leq0.05$ level for $\kpeak \approx 1.2\  h/\mathrm{Mpc}$, but lose sensitivity past $\kpeak \approx 4.5\ h/\mathrm{Mpc}$ (indicated by the red line in Fig.~\ref{figure:constraints}). \lya forest data are sensitive to significantly smaller scales than current CMB data, extending sensitivity to DAOs from \(\sim 1\,h\,\mathrm{Mpc}^{-1}\) to \(\sim 50\,h\,\mathrm{Mpc}^{-1}\). In Ref.~\cite{Barron:2025dys}, UVLF data are used to constrain DAOs, with the same transfer function model that we use.
Bounds on $f$ and $\kpeak$ are derived, but, unlike this work, the DAO amplitude $A$ was not constrained.
While the use of different priors complicates further quantitative comparison, our analysis appears to provide stronger bounds at the smallest scales.

Our baseline model assumes that the same time-varying UVB rates apply at all locations, but this is a known approximation \citep{Chardin:2015uza,Wu:2019sgk,Puchwein:2022wvk}. We thus re-perform the inference using a model of a spatially-inhomogeneous reionization calibrated by radiative transfer simulations \cite{Molaro_2021}, finding that our results do not change (see Appendix \ref{app:systematic_checks}). While our results apply to generic models of interacting dark matter that produce DAOs, we additionally verify that they are directly applicable to models of atomic dark matter. In this case, additional dark radiative cooling can affect astrophysical structures \citep{Roy:2023zar,Roy:2024bcu,Gemmell:2023trd,MandacaruGuerra:2026zqk,Curtin:2020tkm, Curtin:2019lhm, Hippert:2021fch, Armstrong:2023cis,Cabral:2026tjb,Shandera:2018xkn,Gurian:2021qhk,Gurian:2022nbx,Singh:2020wiq,Ryan:2022hku}, but we find that this extra physics has negligible influence on the Lyman-$\alpha$ forest (see Appendix \ref{app:cooling_approx}). This test means that we can directly combine these results with other analyses to probe atomic dark matter, which we will consider in future work.

\textbf{Conclusion and outlook.}
In this work, we present the first Lyman-$\alpha$ forest constraints on DAOs (and indeed any features beyond simple suppression and enhancement) in the linear matter power spectrum using a full statistical analysis that models the flux power spectrum using cosmological hydrodynamical simulations. Given the high cost of simulations and the non-trivial, non-stationary covariance of DAO model parameters, we use a novel combination of deep kernel learning and Bayesian optimization to construct an emulator of the flux power spectra trained on simulations. We use the emulator to perform inference on DAO, cosmological and IGM parameters given the smallest-scale Lyman-$\alpha$ forest data. We find that the Lyman-$\alpha$ forest limits the fraction of dark matter forming DAOs $f \lesssim 0.3$, with stronger bounds $f \lesssim 0.2$ when $\kpeak \sim 1\,h\,\mathrm{Mpc}^{-1}$. These results constitute the first small-scale (\(\kpeak \gtrsim 1\,h\,\mathrm{Mpc}^{-1}\)) cosmological constraints on the DAO amplitude using a full forward model from initial conditions to the IGM. Future work will combine our results with other cosmological datasets~\cite{Barron:2025dys,Barron:2026nks}
to probe concrete physics models like atomic dark matter. We further anticipate that our results can apply to other models like millicharged DM \citep{Dubovsky:2001tr,McDermott:2010pa,Barkana:2018lgd,Boddy:2018kfv}, DM with an electric dipole moment \citep{Sigurdson:2004zp}, DM with massive boson exchange \citep{Dvorkin_2014}, DM interacting with massless sterile neutrinos via a broken dark $U(1)$ gauge symmetry \cite{vandenAarssen:2012vpm,Bringmann:2013vra,Dasgupta:2013zpn}, DM charged under a non-Abelian gauge symmetry \cite{Buen-Abad:2015ova}, and to searches for inflationary potential features \citep{Mergulhao:2023ukp}.

%TC:ignore
\section*{Acknowledgements}
The authors thank Elisa Ferreira for helpful discussions. This work was enabled by computational
resources provided by Compute Canada and the Digital Research Alliance of Canada. The work of ZY, DC and NM was in part supported by Discovery Grants from the Natural Sciences and Engineering Research Council of Canada, the Canada Research Chair program, the Ontario Early Researcher Award, and the University of Toronto McLean Award. KKR is supported by an Ernest Rutherford Fellowship from the UKRI Science and Technology Facilities Council (grant no. ST/Z510191/1). The Dunlap Institute is funded through an endowment established by the David Dunlap family and the University of Toronto. JB acknowledges support from NSF grants PHY-2210533 and PHY-2513893. SR acknowledges support from the Eric and Wendy Schmidt AI in Science Fellowship.

\bibliographystyle{utphys3}
\bibliography{biblio}

\clearpage

\onecolumngrid
\begin{center}
  \textbf{\large Supplemental material for Strongest constraints on dark acoustic oscillations from the Lyman-alpha forest}\\[.2cm]
  \vspace{0.05in}
  {Zhihan Yuan, Caleb Gemmell, Keir K. Rogers, Jared Barron, Sandip Roy, David Curtin and Norman Murray}
\end{center}

\appendix
%%%%%%%%%%%%%%%%%%%%%%%
%%%%%%%%%%%%%%%%%%%%%%%
%%%%%%%%%%%%%%%%%%%%%%%
%%%%%%%%%%%%%%%%%%%%%%%
%%%%%%%%%%%%%%%%%%%%%%%
\section{Dark acoustic oscillation transfer function}
\label{app:transfer_function}

\begin{figure}
    \centering
    \includegraphics[trim={0cm 0cm 0cm 0cm},clip,width=0.8\textwidth]{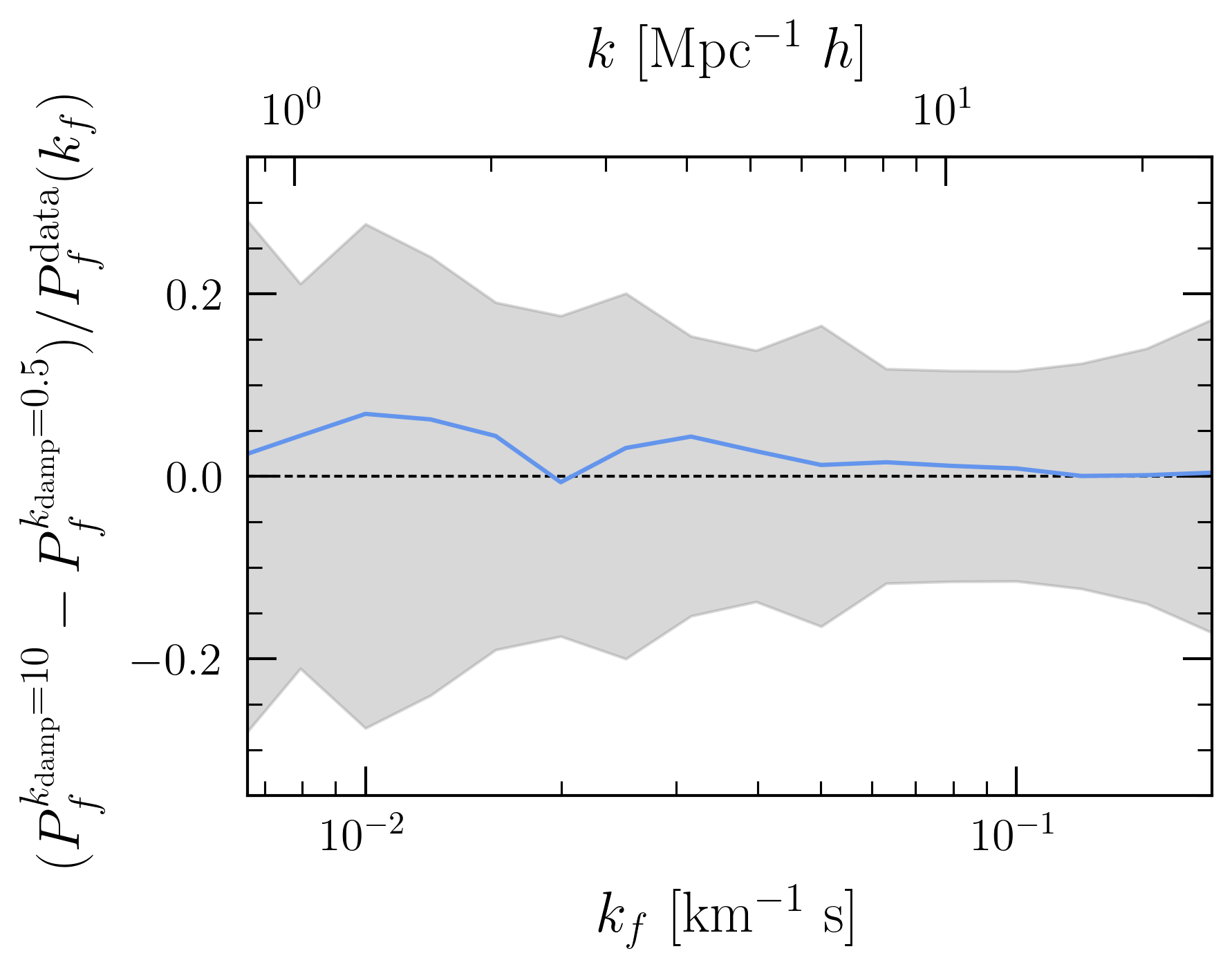}
    \caption{The difference between the flux power spectra at \(z = 4.2\) from a DAO simulation with less damping, $\kdamp = 10$ $\kpeak$, and a DAO simulation with more damping, $\kdamp = 0.5$ $\kpeak$, where $\kpeak = 2$ Mpc$^{-1} h$, normalized by the data. All other model parameters are the same. Grey band depicts the data uncertainty at this redshift. Similar results are found at $z=5$ and $z=4.6$. The simulation with less damping increases the amount of power and is thus a more conservative choice of parameterization when setting constraints. The effect of \(\kdamp\) on the flux power spectrum is in any case weak and within the uncertainty of the data. 
    \label{figure:damp_test}}
\end{figure}

The linear matter power spectrum in the DAO model is defined by (the square of) a transfer function $T(k)$ multiplying the $\lcdm$ linear matter power spectrum such that $P_\mathrm{DAO}(k,z) \equiv T^{2}(k)P_{\lcdm}(k,z)$. The transfer function is the sum of two terms $T=T_{\alpha\beta\gamma\delta} + T_{\mathrm{osc}}$. The first term accounts for the WDM-like power spectrum suppression at high $k$ \citep{Viel:2005qj,Murgia:2017lwo,Hooper:2022byl}: 
\begin{equation}
    T_{\alpha \beta \gamma \delta}(k) \equiv f(1 + (\alpha k)^\beta)^\gamma + (1-f),
    \label{model:fabgd}
\end{equation}
where \(\alpha\), \(\beta\) and \(\gamma\) are free parameters that we will calibrate below. \(f\) is the fraction of the total dark matter that forms DAOs; the remainder is cold dark matter. The second term \(T_\mathrm{osc}\) accounts for the damped DAOs. The oscillations are modeled as sinusoidal with frequency $\omega$ and amplitude \(A\), beginning at wavenumber $k_{\mathrm{start}}$, peaking at wavenumber \(\kpeak\) and damped at wavenumber \(k_\mathrm{damp}\):
\begin{equation}
    \begin{split}
        T_{\mathrm{osc}}(k) \equiv \Theta(k - k_{\mathrm{start}}) \times \left[fA \cos\left(\omega \left(\frac{k}{\kpeak} - 1\right)\right)e^{-\left(\frac{k}{\kdamp}\right)^{2}}\right] \,.
    \end{split}
    \label{model:osc}
\end{equation}

To ensure that our model matches realistic power spectra with strong DAOs, and with future atomic dark matter analyses in mind, we calibrate some of the free parameters by fitting to linear matter power spectra computed using the modified Boltzmann code \texttt{CLASS-aDM} \cite{Bansal:2022qbi}. However, we stress that our parameterization choice nonetheless can be mapped to a wider range of dark sector models with DAOs. We find $\beta = 4.15$, $\gamma = -20$, $\omega=2.083\pi$, $k_{\mathrm{start}}=0.28\kpeak$. The parameter $\alpha$ is chosen such that $T_{\alpha \beta \gamma \delta}(k_{\textrm{start}}) = 0.1 f + (1-f)$, so that the power spectrum suppression matches onto the start of the DAOs:
\begin{equation}
    \alpha = \frac{1}{k_{\mathrm{start}}}\left(0.1^{1/\gamma}-1\right)^{1/\beta} \,.
\end{equation}

After this calibration, we have four free parameters $\{k_{\rm{peak}},k_{\rm{damp}},f,A\}$. To reduce the dimensionality of the parameter space further, we choose to fix $\kdamp = 10 \kpeak$. This decision is motivated by the fact that a larger damping wavenumber (i.e., a less damped oscillation) leads to more power being returned to the power spectrum. Thus, the flux power spectra will be more CDM-like and the constraints we find on DAOs will be necessarily conservative. Fig.~\ref{figure:damp_test} illustrates this fact and also that the flux power spectrum is in any case only weakly sensitive to \(\kdamp\).

%%%%%%%%%%%%%%%%%%%%%%%
%%%%%%%%%%%%%%%%%%%%%%%
%%%%%%%%%%%%%%%%%%%%%%%
%%%%%%%%%%%%%%%%%%%%%%%
%%%%%%%%%%%%%%%%%%%%%%%

\section{Tests of dark radiative cooling in the atomic dark matter model}
\label{app:cooling_approx}

\begin{figure}
    \centering
    \includegraphics[trim={0cm 0cm 0cm 0cm},clip,width=0.8\textwidth]{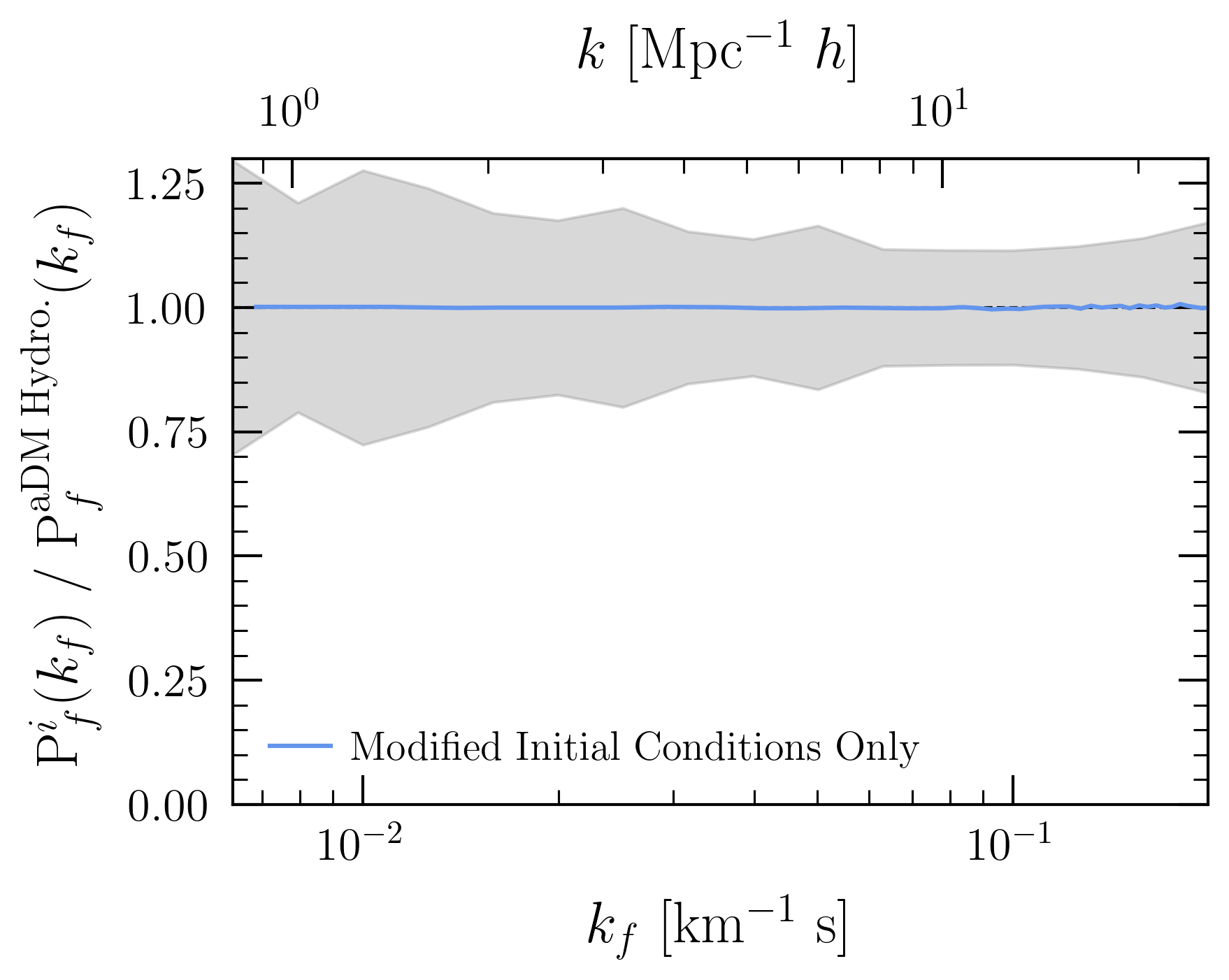}
    \caption{1D flux power spectrum ratio at $z=4.2$ of a (CDM+baryons)-only simulation with modified aDM-like initial conditions relative to a simulation with full aDM hydrodynamics. Grey band depicts the data uncertainty at this redshift. Similar results are found at $z=5$ and $z=4.6$.
    \label{figure:cooling_approx}}
\end{figure}

Dark acoustic oscillations arise in many models of interacting dark matter and dark radiation, but one model of particular interest is atomic dark matter (aDM)~\cite{Kaplan:2009de}. In addition to forming DAOs, aDM can undergo radiative and collisional cooling, which can also affect astrophysical structure~\cite{Fan:2013tia,Fan:2013yva}. While DAOs occur at much earlier redshifts and are accounted for in \texttt{CLASS}, the cooling effects become important at much later redshifts and require a specialized version of \texttt{GIZMO}~\cite{Roy:2023zar} to incorporate. In this implementation, aDM is treated hydrodynamically as a gas, similar to baryons in the original version, making it more computationally expensive. For the DAO results we present here to apply to models of aDM, we must verify that aDM cooling has a negligible effect on the Lyman-$\alpha$ forest.

We hypothesize that since the Lyman-$\alpha$ forest signal derives from the IGM, and that aDM cooling occurs in the centers of halos, the cooling will have little effect on the flux power spectra. We explicitly test this approximation by considering an aDM parameter point expected to cool significantly, with a large dark matter fraction in aDM ($f = 0.9$), a large dark temperature ratio ($\xi = 0.5$), and a small $\beta_{\rm{cool}}$ ($\sim0.001$)\footnote{Lower $\beta_{\rm cool}$ signifies more efficient dissipation.} as defined in Ref.~\cite{Roy:2024bcu}. At this parameter point, we run one simulation with the full aDM hydrodynamics, and one CDM+baryons-only simulation but where we initialize the CDM particles to have the same matter power spectrum as the aDM simulation at $z=99$. Fig.~\ref{figure:cooling_approx} shows that we indeed find no appreciable effect on the flux power spectrum from aDM cooling.

However, while aDM cooling does not have a direct effect on the Lyman-$\alpha$ forest through the distribution of matter in the IGM, cooling in halos could have a large impact on star formation rates and thus the UV background photons that heat and ionize the IGM. Since we already consider a large range of IGM histories by varying the input simulation parameters, $\{H_A, H_S,z_{\rm{rei}},T_{\rm{rei}}\}$, we anticipate that this aDM effect would be degenerate with the output IGM parameters we already consider.

\section{Tests of \texttt{QuickLymanAlpha} flag and feedback models}
\label{app:fast_flag}

\begin{figure}
    \centering
    \includegraphics[trim={0cm 0cm 0cm 0cm},clip,width=0.8\textwidth]{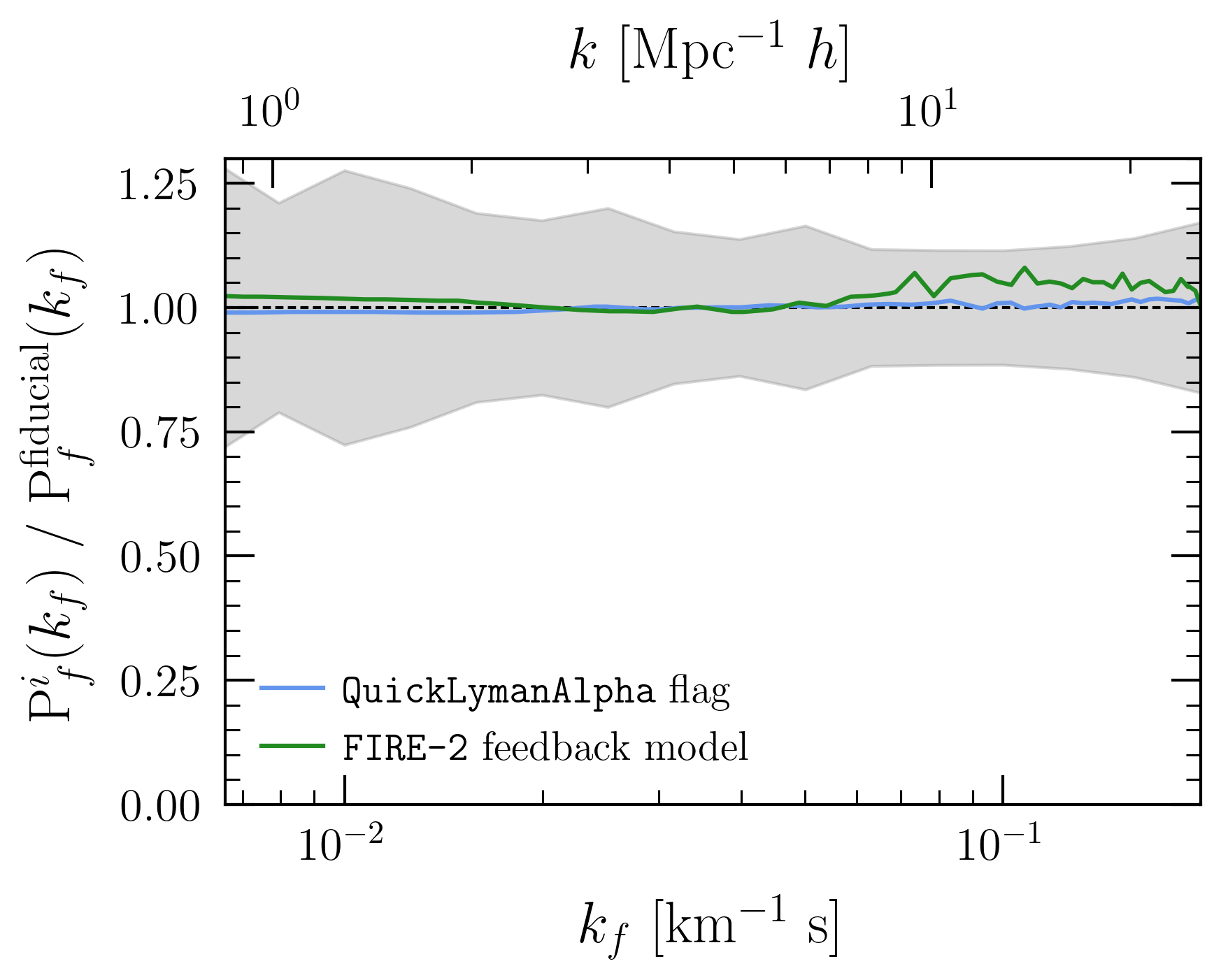}
    \caption{1D flux power spectra ratios at $z=4.2$ relative to a simulation with default \texttt{GIZMO} feedback flags. We compare a \texttt{QuickLymanAlpha} simulation (blue) and a \texttt{FIRE-2} feedback simulation (green). Grey band depicts the data uncertainty at this redshift. Similar results are found at $z=5$ and $z=4.6$.
    \label{figure:fast_flag}}
\end{figure}

Previous Lyman-$\alpha$ forest studies have used a simplified star formation criterion to speed up simulations while having negligible impact on the flux power spectra \cite{Viel:2004bf}. Usually referred to as the \texttt{QuickLymanAlpha} flag, the criterion converts gas particles at overdensities $>$ 1000 and with temperatures $<$ $10^5$ K into collisionless star particles. Fig.~\ref{figure:fast_flag} confirms that there is no significant difference between the \texttt{GIZMO} simulation with and without the \texttt{QuickLymanAlpha} implementation (\textit{blue line}). Further, studies have shown that active galactic nuclei (AGN) feedback can have a marginal effect on the 1D flux power spectra \cite{Chabanier:2020uuh,2024MNRAS.532.4876D,Tillman:2024pim}. To explore this effect, we compare the effect of the \texttt{FIRE-2} feedback model \cite{Hopkins:2017ycn} to the default setting (\textit{green line}). While the \texttt{FIRE-2} module results in a modest $\sim(5-10)\%$ difference at smaller scales, it is within the error bars of the data. Thus, for the sake of computational speed, we opt to use the \texttt{QuickLymanAlpha} flag and neglect using \texttt{FIRE-2}.

\begin{figure}
    \centering
    \includegraphics[trim={0cm 0cm 0cm 0cm},clip,width=0.8\textwidth]{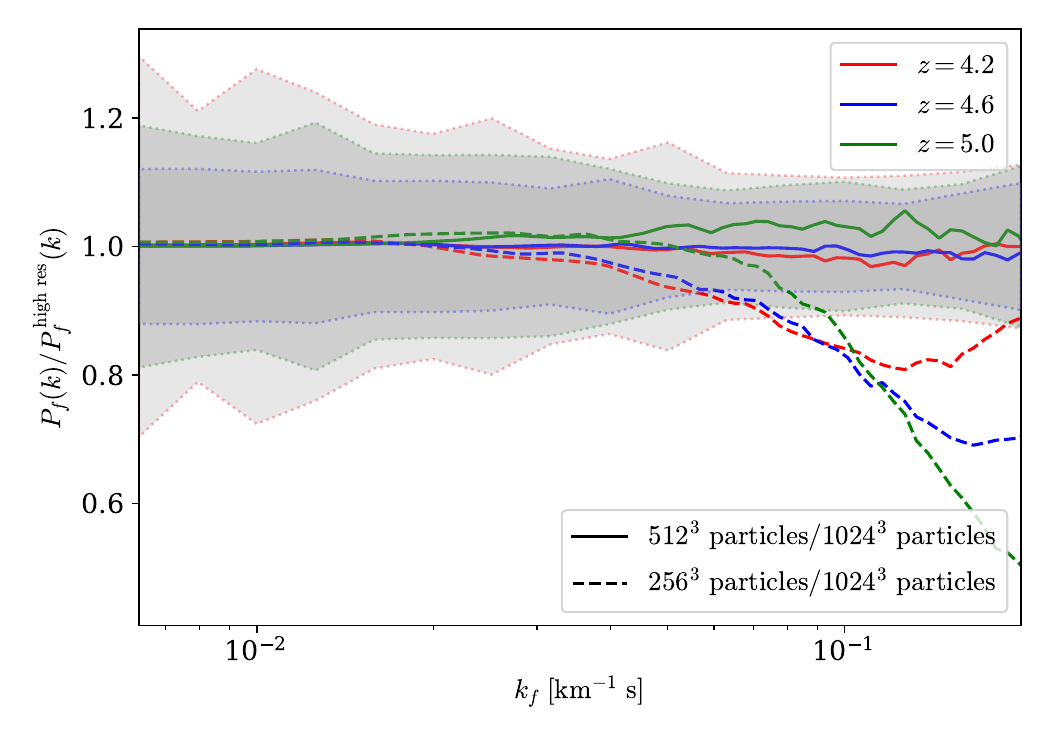}
    \caption{A test of convergence in the simulated flux power spectra with respect to the number of simulation particles in a fixed-size box ($10\, h^{-1}\,\mathrm{Mpc}$). The solid lines show the ratio between the flux power spectra of a simulation with $2 \times 512^3$ particles and a simulation with $2 \times 1024^3$ particles. The dashed lines show the ratio between a simulation with $2 \times 256^3$ particles and a simulation with $2 \times 1024^3$ particles. The color indicates the redshift. The gray shaded regions indicate the data uncertainties (outlined by red, blue and green according to their redshift).}
    \label{figure:res}
\end{figure}

\section{Numerical convergence tests}
\label{app: convergence}

To ensure that we have chosen a sufficient number of simulation particles, we run test simulations with different particle resolutions for a fixed box size. We run this test at a parameter point with strong matter power spectrum suppression, so that any numerical effects from small-scale fragmentation would be most apparent. Fig.~\ref{figure:res} shows the convergence tests for particle resolution. The ratio between the flux power spectrum from the $512^3$ particle simulation to that from the $1024^3$ particle simulation remains within the data error.

\begin{figure}
    \centering
    \includegraphics[trim={0cm 0cm 0cm 0cm},clip,width=0.8\textwidth]{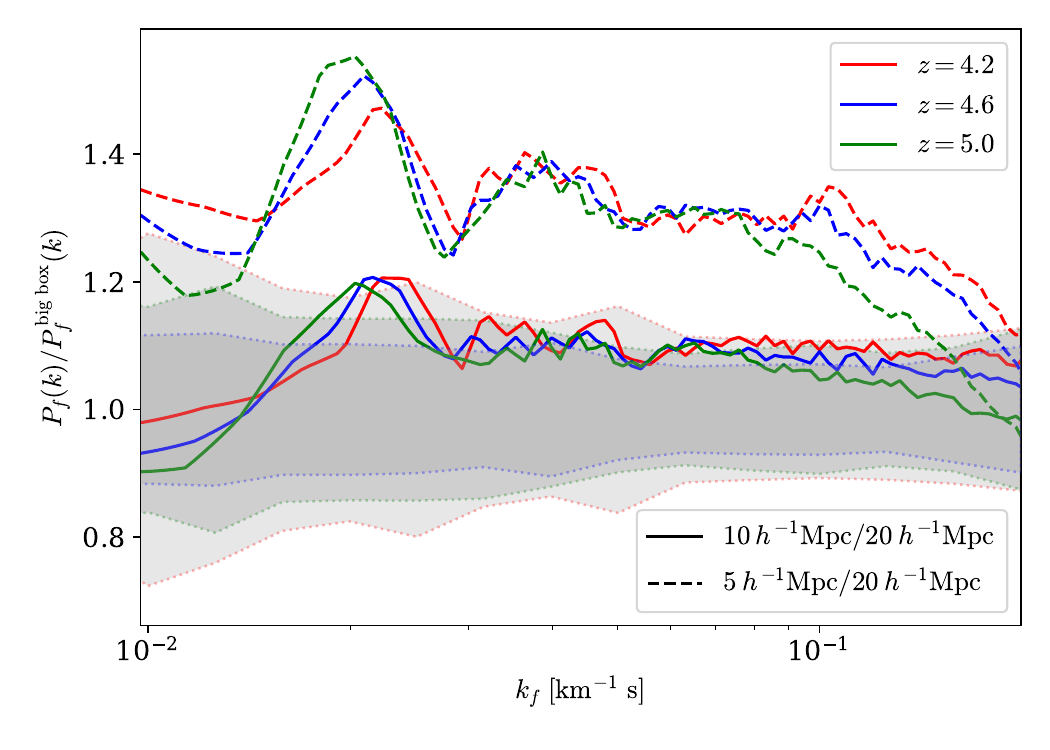}
    \caption{A test of convergence in the simulated flux power spectra with respect to the simulation volume with fixed particle mass. The solid lines show the ratio between the flux power spectra of a simulation with volume $10\,h^{-1}\,\mathrm{Mpc}$ and a simulation with volume $20\,h^{-1}\,\mathrm{Mpc}$. The dashed lines show the ratio between a simulation with volume $5\,h^{-1}\,\mathrm{Mpc}$ and a simulation with volume $20\,h^{-1}\,\mathrm{Mpc}$. The color indicates the redshift. The gray shaded regions indicate the data uncertainties (outlined by red, blue and green according to their redshift).}
    \label{figure:box_effect}
\end{figure}

In optically-thin Lyman-$\alpha$ forest simulations, the finite box size introduces a systematic bias in the predicted flux power spectra due to the absence of long-wavelength modes and their nonlinear coupling to smaller scales \cite{Tytler_2009,Luki__2014}. The missing large-scale modes lead to smaller bulk flow velocities and less shock heating in the IGM, thereby yielding systematically colder gas. These effects reduce thermal broadening and Jeans smoothing, enhancing small-scale flux power. The power enhancement is expected to increase monotonically with smaller box size, as seen in Refs.~\cite{Tytler_2009,Luki__2014}. We confirm this monotonicity with simulations with box sizes of $5\,h^{-1}$ Mpc, $10\,h^{-1}$ Mpc, and $20\,h^{-1}$ Mpc, as shown in Fig.~\ref{figure:box_effect}. To reduce the effects of sample variance, we average over ten simulations with different random seeds for the $5\,h^{-1}$ Mpc and $10\,h^{-1}$ Mpc boxes. We correct for the box size effect by rescaling the flux power spectra of our $10\,h^{-1}$ Mpc simulations to match the larger boxes, as described in the main text.

%%%%%%%%%%%%%%%%%%%%%%%
%%%%%%%%%%%%%%%%%%%%%%%
%%%%%%%%%%%%%%%%%%%%%%%
%%%%%%%%%%%%%%%%%%%%%%%
%%%%%%%%%%%%%%%%%%%%%%%
\section{Deep kernel learning emulator and Bayesian optimization}
\label{app: emulator}

\begin{figure}
    \centering
    \includegraphics[trim={0cm 0cm 0cm 0cm},clip,width=0.8\textwidth]{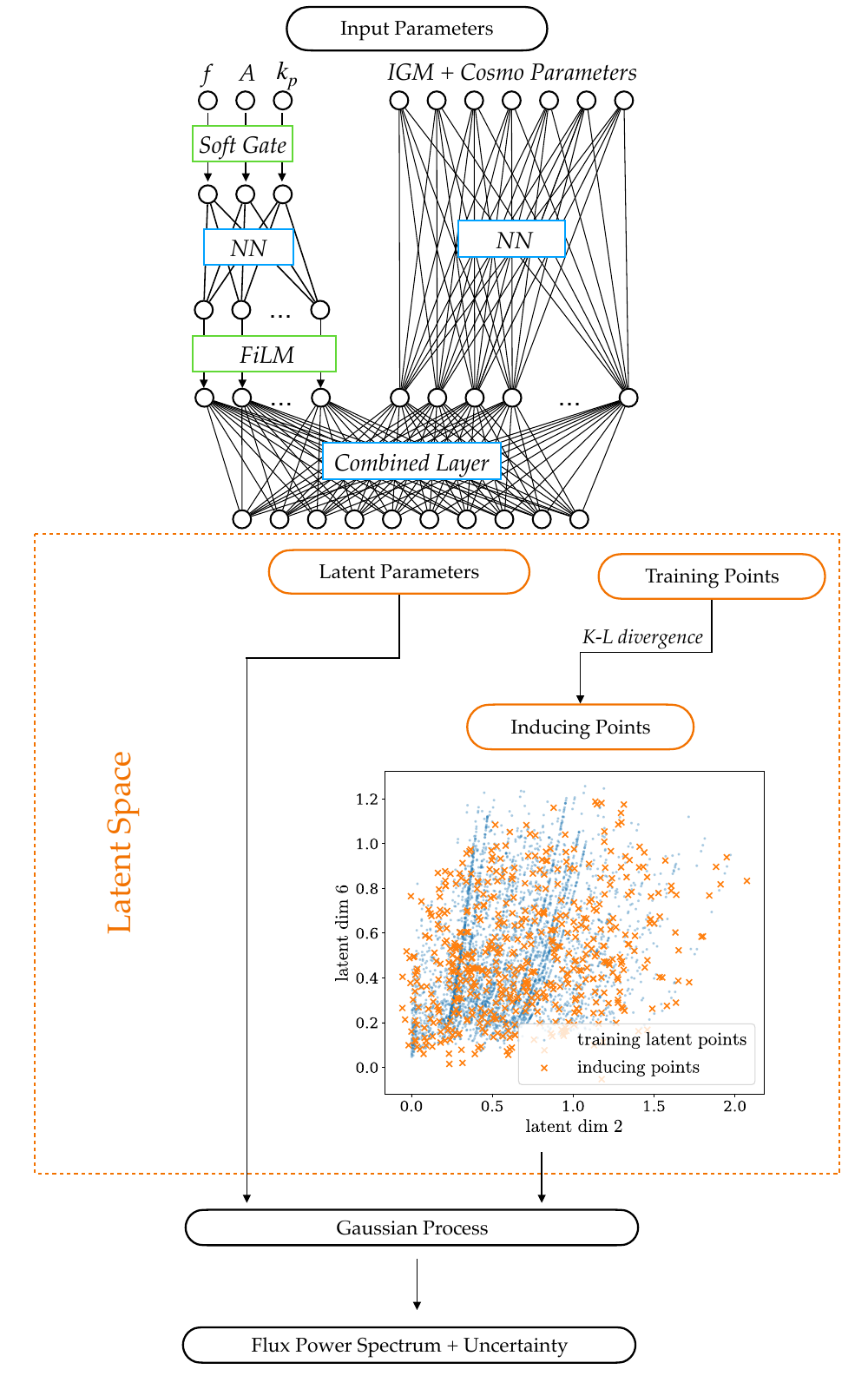}
    \caption{Deep kernel learning emulator flow chart with the feature extractor at the top and GP at the bottom. \label{figure:dkl_flowchart}}
\end{figure}

Gaussian processes (GPs) have been used previously to emulate the Lyman-$\alpha$ forest flux power spectrum \cite{Rogers:2018smb,Bird:2018efe,Rogers:2020ltq,Rogers:2020cup,Pedersen:2020kaw,Rogers_2022,Bird:2023evb}. A GP model describes a prior distribution over functions $g(x)$ which can be specified by a mean function $m(x)$ and a covariance (kernel) function $\mathrm{ker}(x,x')$. In our context here, \(x\) is the DAO, cosmological and IGM parameter vector and \(g\)  is the mapping to the flux power spectrum vector evaluated at fixed values of the velocity wavenumber \(k_\mathrm{f}\) and redshift \(z\). We are free to select different kernel functions to encode desired properties of the model. If we condition this Gaussian process prior on a training dataset of simulations, we obtain a posterior distribution that allows us to make predictions of the flux power spectrum at other parameter points. Given a simulated training dataset of pairs of parameter points and flux power spectra ${X,Y}$ with noise $\sigma_n$, the covariance matrix becomes

\begin{equation}
    K_Y = \mathrm{ker}(X,X)+\sigma_n^2I,
\end{equation}
where $I$ is the identity matrix. If we want to make a prediction at parameter point $x'$, the joint prior distribution is 
\begin{equation}
    \left[\begin{array}{c}
\mathbf{Y} \\
g(x')
\end{array}\right] \sim \mathcal{N}\left(\left[\begin{array}{c}
m \\
m'
\end{array}\right],\left[\begin{array}{cc}
K_Y & \mathrm{ker}(X,x') \\
\mathrm{ker}(X,x')^T & \mathrm{ker}(x', x')
\end{array}\right]\right).
\end{equation}
Conditioning on the simulated training data, we get a Gaussian-distributed predictive posterior $p(g(x')|Y)$ given by the posterior mean
\begin{equation}
    \mu'=m'+\mathrm{ker}(X,x')^TK_Y^{-1}(Y-m)
\end{equation}
and posterior variance
\begin{equation}
    \sigma'^2=\mathrm{ker}(x',x')-\mathrm{ker}(X,x')^TK_Y^{-1}\mathrm{ker}(X,x').
    \label{eq:GP_variance}
\end{equation}
This posterior is the emulator prediction for the flux power spectrum at a new DAO, cosmological and IGM parameter point.

However, despite its previous success for Lyman-$\alpha$ forest flux power spectrum emulation, a traditional GP emulator is not sufficient to describe the full complexity of the DAO effects on the Lyman-$\alpha$ forest flux power spectra, even by considering different covariance kernel function choices. Specifically, the covariance of our parameter space is highly non-stationary. As \(f \rightarrow 0\), the effects of \(\kpeak\) and \(A\) on the flux power vanish, i.e., their covariance changes. There do exist non-stationary covariance kernels such as the Gibbs kernel but we leave detailed comparisons to future work. 

To represent fully these behaviors, we turn to deep kernel learning (DKL) \cite{wilson16}, where a small neural network works as a feature extractor and maps the original parameter space to a latent parameter space, where we enforce stationarity. We then use a GP to emulate the flux power spectrum given this new latent parameter space. Other emulation strategies use a neural network to map directly from parameters to simulation outputs like the power spectrum \citep[e.g.,][]{SpurioMancini:2021ppk,Piras:2023aub,Cabayol-Garcia:2023ygj}. However, these do not typically provide a global uncertainty quantification (UQ). Bayesian neural networks \citep[e.g.,][]{Mancarella:2020jyu,Lemos:2022kua} can combine deep learning emulation with global UQ, but also typically suffer from underdetermined hyperparameters. UQ is important in our case as we will use posterior predictive uncertainties to design the simulation training set by active learning (Bayesian optimization).

Figure \ref{figure:dkl_flowchart} shows a flow chart that illustrates the architecture of our DKL emulator. The feature extractor network is at the top. Its architecture is physically informed. The neural network is split into two branches: one for DAO transfer function parameters and the other for cosmological and IGM parameters. In the transfer function branch, we first multiply a soft gate function $G(f)$ to $k_{\rm{peak}}$ and $A$ to enforce the $\Lambda$CDM behavior at $f=0$, where $k_{\rm{peak}}$ and $A$ do not change the transfer function:

\begin{equation}
    G(f)=1 - e^{-(f/\tau)^p},
\end{equation}
where hyperparameters $\tau=0.02$ and $p=3$ are fixed to ensure that the gate function turns on smoothly as $f$ increases.  The network is then followed by two linear layers with widths 16 and 8, respectively, each followed by a Softplus activation. An additional feature-wise linear modulation (FiLM) layer \cite{perez_2017} is applied to the latent representation of the transfer function branch, allowing the learned features to be conditioned continuously on the DAO parameter $f$. The FiLM layer thus applies a transformation
\begin{equation}
    h' = \gamma(f)h+\beta(f),
\end{equation}
where $h$ and $h'$ are the layer before and after modulation, $\gamma$ and $\beta$ are functions of $f$ learned through a small network. They are optimized such that $h'=h$ when $f=0$, corresponding to the desired $\Lambda$CDM behavior. This modulation thus smoothly encodes the physical behavior of $f$ so that the total network does not have to ``jump" between drastically different behaviors when approaching the \(\Lambda\)CDM limit. The cosmological and IGM parameter branch consists of a single 16-dimensional fully connected layer. The 8-dimensional physics representation and the 16-dimensional nuisance representation are concatenated into a 24-dimensional feature vector. This vector is subsequently transformed through a linear layer followed by a Softplus activation, yielding 10 output latent dimensions, which matches the input dimensionality.\footnote{Although the full model has 18 parameters, we train a separate emulator for each redshift bin, leading to ten dimensions per emulator.}

For the second stage of the DKL (see bottom of Fig.~\ref{figure:dkl_flowchart}), we use a Gaussian process emulator that now takes pairs of parameters in the latent space and corresponding simulated flux power spectra and then outputs predicted flux power spectra at new parameter points with uncertainties quantified. We use a variational Gaussian process, also called a sparse Gaussian process \cite{rasmussen_williams_2008,titsias09a,leibfried2022tutorialsparsegaussianprocesses}. The main advantage of the variational GP over the traditional GP described above is its scalability with large training sets. In this work, the training set has 4130 points (including accounting for the mean flux rescaling).  The variational GP has much smaller memory requirements and faster evaluation time since it avoids inverting the matrix \(K_Y\) for the total training set. This scalability is necessary for DKL, because the memory requirement and evaluation time quickly becomes substantial if each training step requires the evaluation of a deep neural network and an exact GP for the full dataset. Variational GP thus naturally supports batch training, which makes it much more compatible with neural networks.

\begin{figure}
    \centering
    \includegraphics[trim={0cm 0cm 0cm 0cm},clip,width=0.9\textwidth]{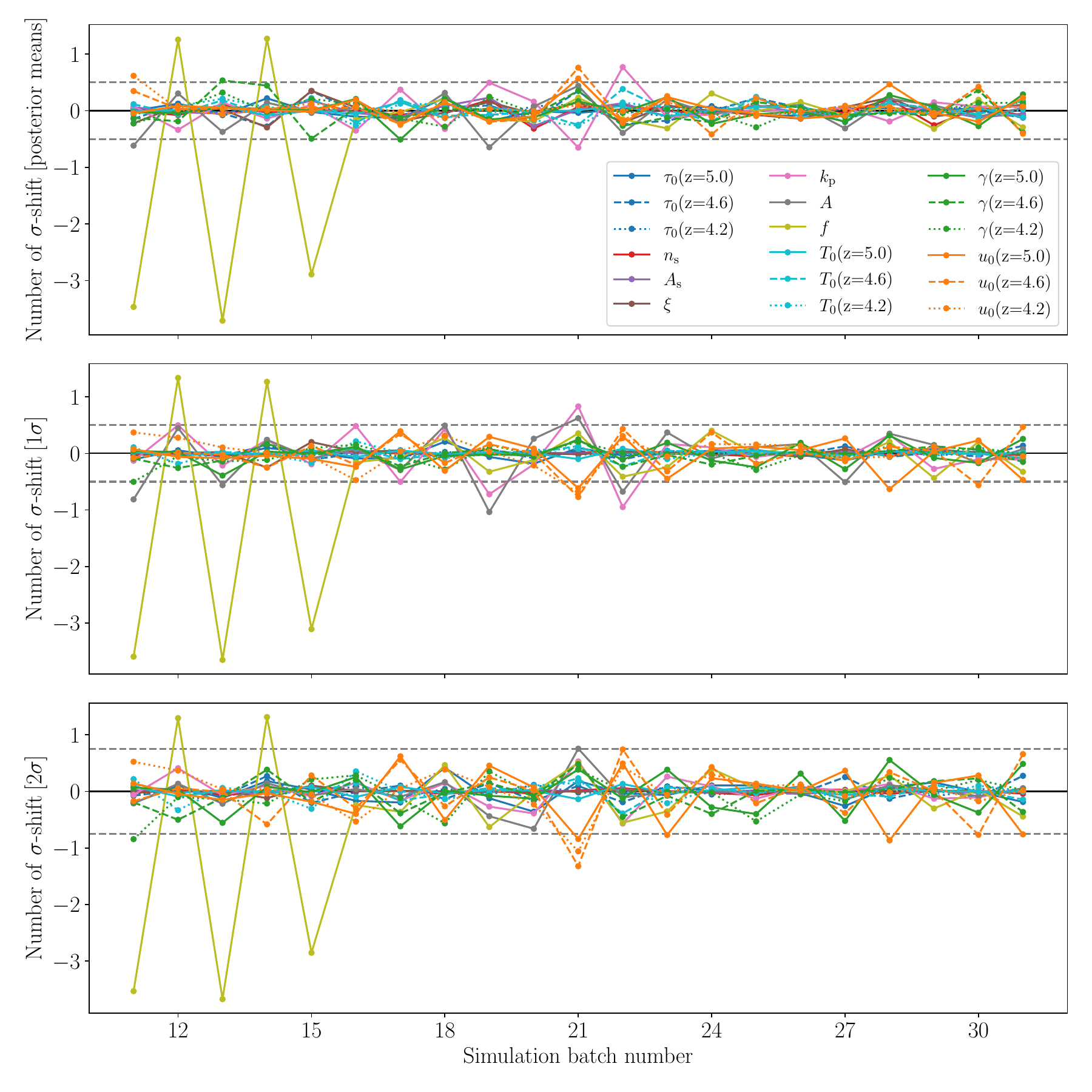}
    \caption{The convergence of summary statistics of the posterior distribution of DAO, cosmological and IGM parameters given flux power spectrum data using the Bayesian-optimized emulator. \textit{From top to bottom}, we show the number of sigma shift (defined by the marginalized posteriors at a given optimization epoch) between emulator iterations for the 1D marginalized posterior means and 1$\sigma$ and 2$\sigma$ constraints. Each colored line shows the convergence for each model parameter.}
    \label{figure:convergence}
\end{figure}

We now describe the variational GP approach. For a training dataset, let $X'$ be the set of $n$ parameter points and $Y$ be the noisy evaluation of some latent function $g(X')$. For clarity, as above, \(X'\) is the set of training DAO, cosmological and IGM parameters, now mapped to the new latent space, \(g\) is the mapping to simulated flux power spectra, at fixed values of \(k_f\) and \(z\), and \(Y\) is the set of flux power spectra evaluated at the training points. The noise here refers to sample variance in the simulations. As in the case of the exact GP emulator described above, we introduce a Gaussian process prior to $g$, specified by a mean function and a covariance kernel, and we want to obtain the predictive posterior.
For the variational GP, $m$ ``inducing points'' are chosen in the same (latent) space of $X'$, but at different points than the training set. Let the set of inducing points be $Z$. The flux power spectrum values $u \equiv g(Z)$ evaluated at each inducing point are treated as a compact set of variables that summarizes the behavior of the full GP. Unlike the training targets, the inducing values are not directly observed. Instead, their posterior distribution is learned during training. The exact predictive posterior is thus approximated as
\begin{equation}
    p(g|X',Y)\approx q(g)=\int p(g|u)q(u) \dif u,
\end{equation}
where $q(u)$ is the prior distribution for the inducing points. On the covariance level, this means we can approximate the true covariance $K_{n n}\approx K_{n m} K_{m m}^{-1} K_{m n}$
where $K_{m m}$ is the covariance matrix on the inducing points, and $K_{n m}$ is the cross-covariance matrix between training and inducing points.
The computation now scales with $\mathcal{O}\left(nm^2\right)$ instead of $\mathcal{O}\left(n^3\right)$. In our case, we have 4130 training points and only 128 inducing points.

\begin{figure}
    \centering
    \includegraphics[trim={0cm 0cm 0cm 0cm},clip,width=\textwidth]{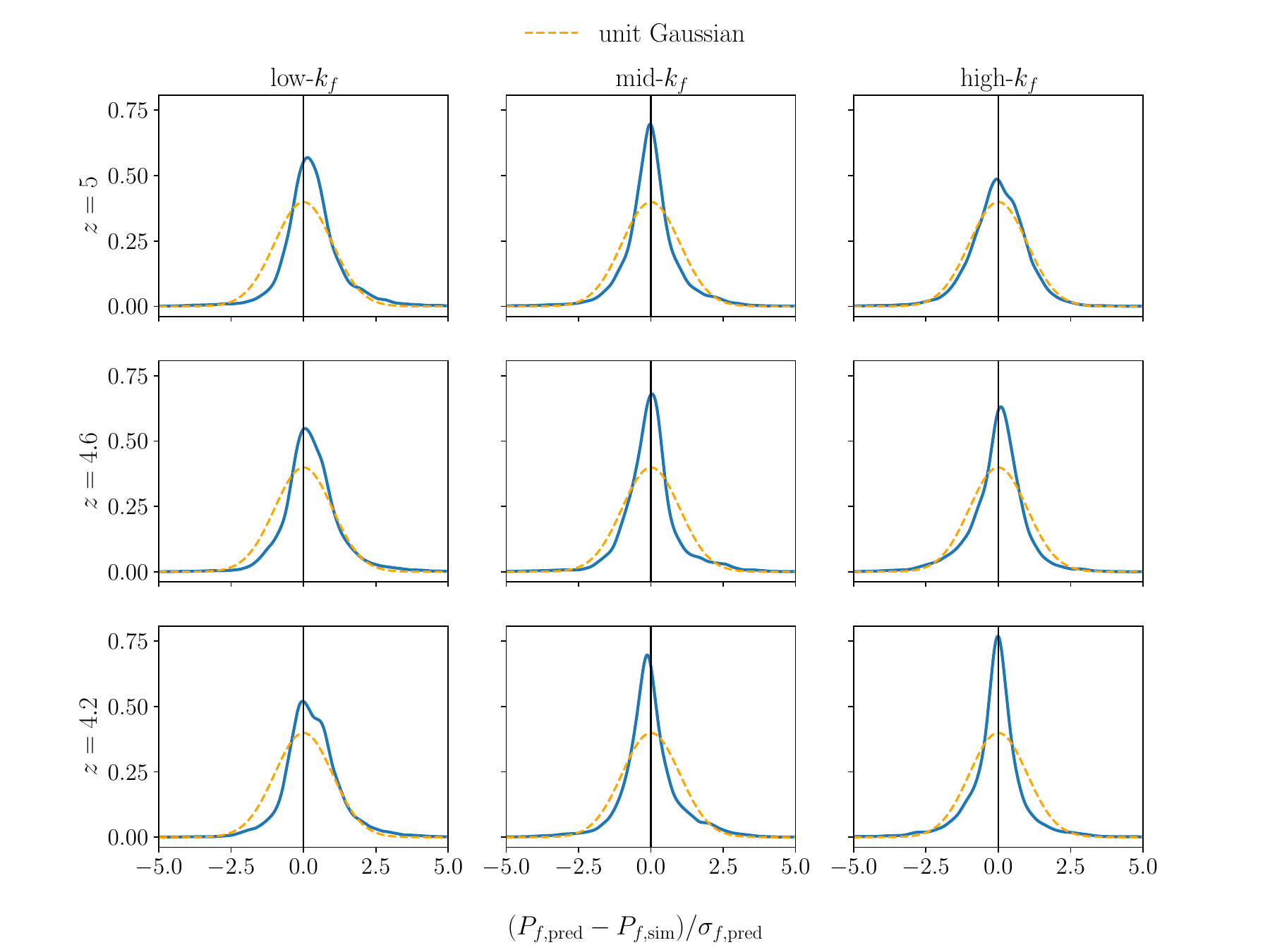}
    \caption{Leave-one-out cross-validation emulator test. Distribution of the ratio of empirical empirical error \(P_{f,\mathrm{pred}} - P_{f,\mathrm{sim}}\) to estimated emulator error \(\sigma_{f,\mathrm{pred}}\) from the Gaussian process for all training simulations at all three redshifts (\textit{from top to bottom}) and for low-\(k_f\), mid-\(k_f\), and high-\(k_f\) wavenumber bins (\textit{from left to right}; solid blue lines). Orange dashed lines show a unit Gaussian distribution. The leave-one-out distributions are consistently more peaked than a unit Gaussian, indicating that the emulator is mildly underfitting, i.e., that the estimated uncertainty on the emulated flux power spectrum is larger than the true uncertainty. This underfitting will lead to conservative bounds on model parameters after propagation to the likelihood function. See main text for more details. }
    \label{figure:kde1}
\end{figure}

The inducing points are selected by minimizing the Kullback-Leibler (KL) divergence between the approximate posterior \(q\) using the inducing points and the exact posterior \(p\) using the training points:
\begin{equation}
    KL(q(g)||p(g|X',Y))=\int \dif f q(g) \log {\frac{q(g)}{p(g|X',Y)}}.
\label{eq:KL}
\end{equation}
However, we want to avoid directly computing the exact posterior \(p\), which is the motivation of using the variational GP. To minimize the KL divergence without such an explicit computation, we use Bayes' theorem such that
\begin{equation}
\log{p(g|X',Y)}=\log{p(Y|g,X')}+\log{p(g)}-\log{p(Y|X')}.
\label{eq:bayes}
\end{equation}
It follows by combining Eqs. (\ref{eq:KL}) and (\ref{eq:bayes}) that 
\begin{equation}
    \log{p(Y|X')}-KL(q(g)||p(g|Y))=\int \dif g q(g)\log{p(Y|g)}-KL(q(g)||p(g)) \equiv \mathrm{ELBO}.
\end{equation}
We define the above to be the evidence lower bound (ELBO), which we maximize during training. From the left hand side, we can see that maximizing ELBO simultaneously maximizes the marginal likelihood \(p(Y|X')\), which improves the fit to the simulated data, and minimizes the KL divergence to the true posterior, which optimizes inducing point selection. The right hand side is directly used to compute ELBO during each training step. In our architechture, the feature extractor and the GP are jointly trained for 150 epochs before we freeze the neural network hyperparameters and we then fine-tune the GP, using the ELBO \cite{leibfried2022tutorialsparsegaussianprocesses} as the loss function. The model is trained using the Adam optimizer. We illustrate the projection of training and inducing points onto two of the latent parameters in Fig.~\ref{figure:dkl_flowchart}. As expected, the density of inducing points roughly traces the density of training points. The streaks in the latent space projection come from the mean flux rescalings that more densely sample the \(\tau_0\) dimensions.

\begin{figure}
    \centering
    \includegraphics[trim={0cm 0cm 0cm 0cm},clip,width=\textwidth]{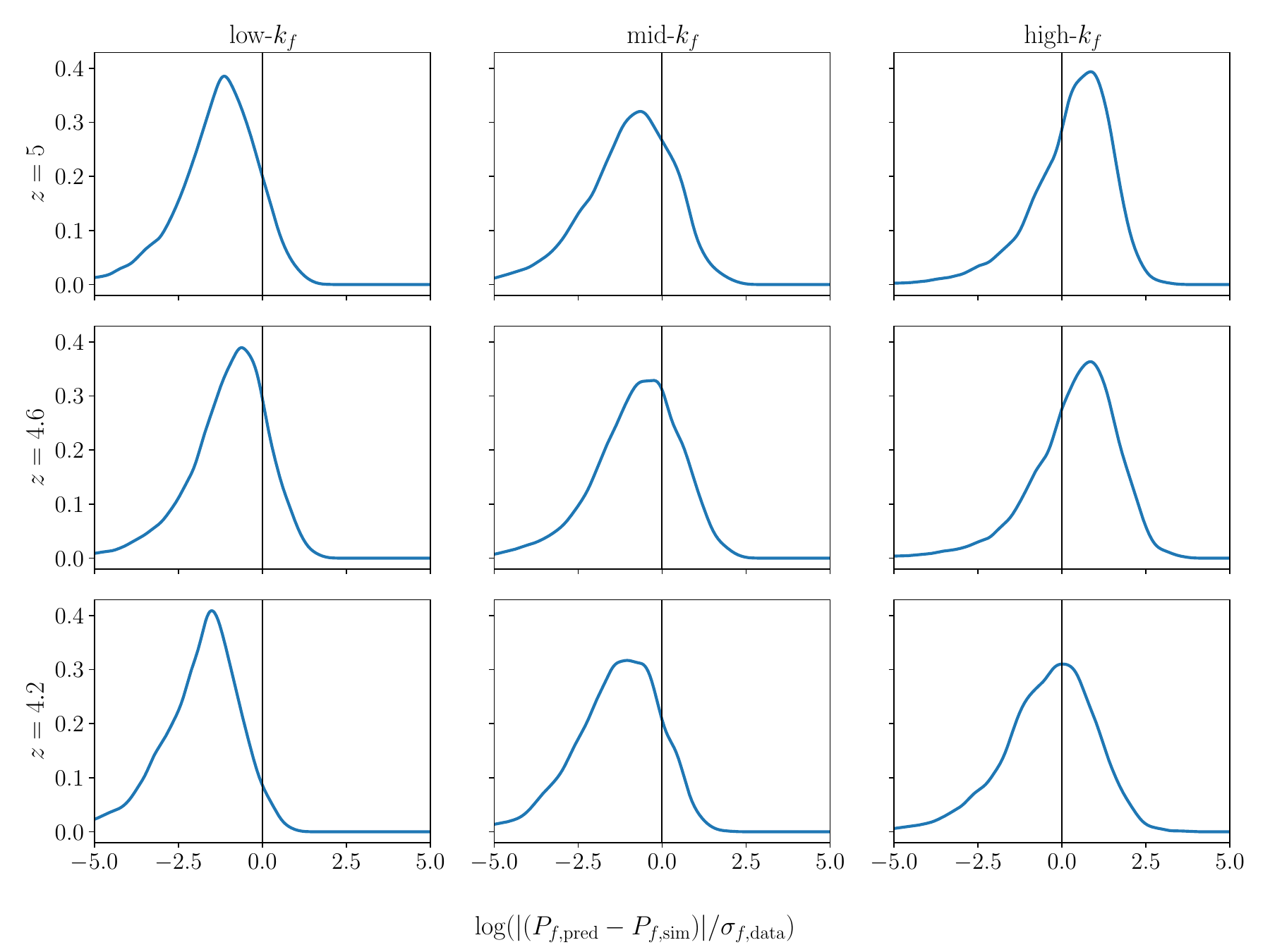}
    \caption{Distribution of the logarithm of the ratio of empirical empirical error \(|P_{f,\mathrm{pred}} - P_{f,\mathrm{sim}}|\) (estimated by leave-one-out cross-validation) to data error \(\sigma_{f,\mathrm{data}}\) for all training simulations at all three redshifts (\textit{from top to bottom}) and for low-\(k_f\), mid-\(k_f\), and high-\(k_f\) wavenumber bins (\textit{from left to right}; solid blue lines). For low-\(k_f\) and mid-\(k_f\) bins, the emulator uncertainty is usually much less than the data uncertainty. For the high-\(k_f\) bins, the emulator uncertainty is comparable in magnitude; this uncertainty is propagated to the likelihood function leading to conservative model parameter bounds. }
    \label{figure:kde2}
\end{figure}

In summary (see also Fig.~\ref{figure:dkl_flowchart}), our input parameters are fed into a feature extractor that consists of a neural network split into two branches, where the physically informed branch contains a soft gate and a layer of FiLM modulation to express known parameter dependencies. The feature extractor produces latent parameters, which are then fed into a variational GP model. The GP model consists of a posterior distribution approximated using a set of inducing points in the latent parameter space, which are selected during training to maximally represent training data. We use this approximate posterior distribution to predict the flux power spectrum. We iteratively expand the training set using the Bayesian optimization procedure described in Ref.~\cite{Rogers:2018smb}. At each Bayesian optimization step, we re-train the DKL emulator using the procedure described above. In practice, we train three emulators, one for the flux power spectrum at each redshift that we consider.

\begin{figure}
    \centering
    \includegraphics[trim={0cm 0cm 0cm 0cm},clip,width=0.8\textwidth]{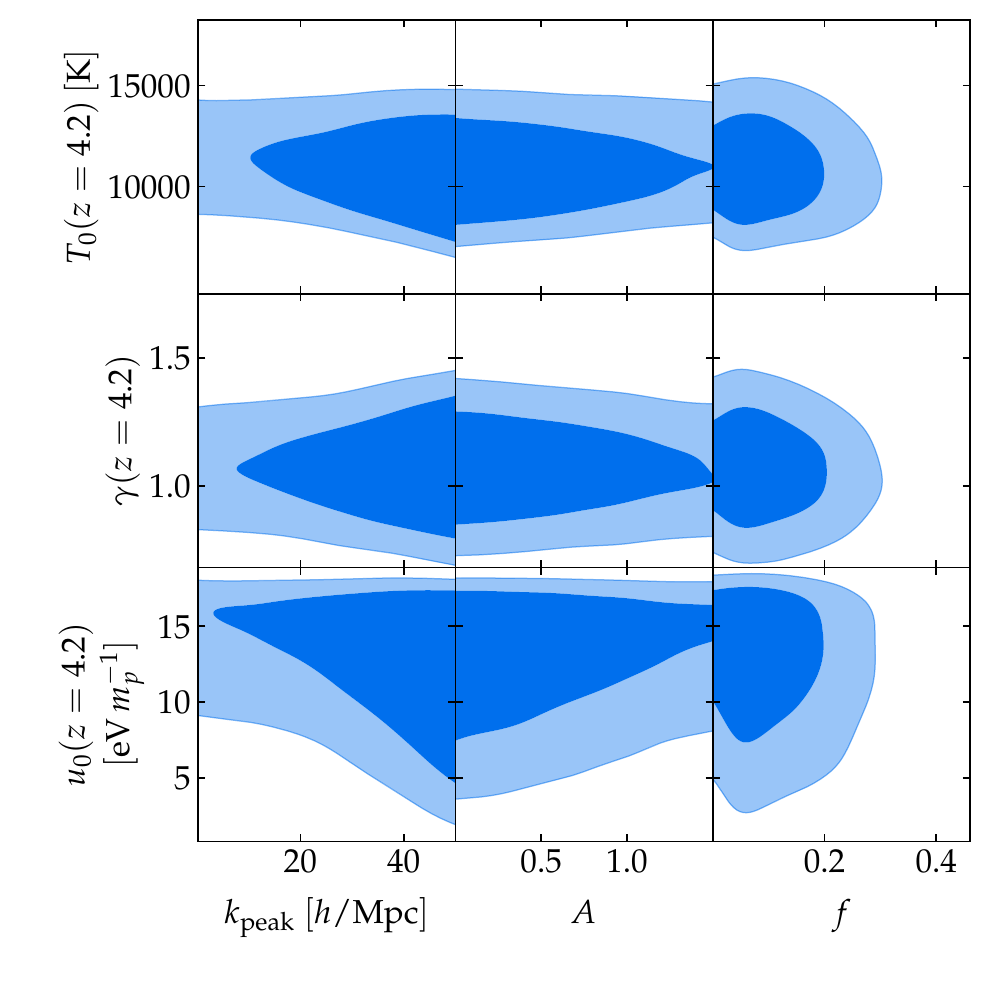}
    \caption{As Fig.~\ref{figure:constraints}, but showing degeneracies between DAO transfer function and IGM parameters at \(z=4.2\). Similar results are found at \(z=5\) and \(z=4.6\).}
    \label{figure:degen}
\end{figure}

Figure \ref{figure:convergence} shows the DAO, cosmological and IGM model parameters' shift by number of sigmas between each Bayesian optimization training set iteration. We start with 68 simulations sampled by a Latin hypercube in the parameter space (680 training points in total after mean flux rescaling post-processing). Bayesian optimization adds training simulations by a balance of selecting points that have high emulator prediction uncertainty (exploration) and points that have high posterior probability given the true data (exploitation). The simulation batch size that we add at each step varies from 10 to 5, with an exception in the last few fine-tuning batches, which only contain 2-3 simulations each. The final training set contains 413 simulations (4130 training points in total), now clustered in regions of high posterior probability given the flux power spectrum data. We observe that the variation between the posteriors given successive training set iterations is small by the end, indicating an emulator that is stable with respect to the training set design. Indeed, most model parameters have already converged by batch 16, but the DAO parameters and the IGM parameters $u_0(z=z_i)$ are not stable until later.

Figures \ref{figure:kde1} and \ref{figure:kde2} show kernel density estimates of the distributions of the ratio of empirical emulator error to estimated emulator error and data error, respectively. Empirical emulator error is defined as the leave-one-out cross-validation error, i.e., re-training the emulator after leaving out each of the training simulations\footnote{In practice, we leave out all ten mean flux rescalings per simulation, making this in effect leave-\textit{ten}-out cross-validation.} and comparing the flux power prediction \(P_\mathrm{f,pred}\) to the true simulated flux power \(P_\mathrm{f,sim}\). Estimated emulator error is the standard deviation of the posterior predictive distribution of the flux power spectrum given the Gaussian process emulator in the leave-one-out cross-validation setting, i.e., \(\sigma\) as defined in Eq.~\eqref{eq:GP_variance}. Data error is as estimated in Ref.~\cite{Boera:2018vzq}. We find that the emulator error is consistently overestimated with respect to the true error at all wavenumbers and redshifts (the blue distributions are more peaked than the orange), i.e., the emulator is underfitting. After propagation of the emulator uncertainty into the data likelihood function, this will lead to more conservative model parameter constraints. We find that for the low-\(k_\mathrm{f}\) and mid-\(k_\mathrm{f}\) bins, the emulator error is usually much less than the data uncertainty. For the high-\(k_\mathrm{f}\) bins, the uncertainties are comparable, which, again after considering propagation to the likelihood, will lead to more conservative bounds. This result is consistent with previous flux power spectrum emulators \citep[e.g.,][]{Chaves-Montero:2026hqd}, where emulator uncertainties contribute non-negligibly to the total error budget. We leave to future work an investigation into strategies to reduce further this part of the error budget, e.g., by using multi-fidelity emulators \citep[e.g.,][]{Fernandez:2022kxq,Ho:2023alj} to distribute training simulations even more optimally.

\section{Degeneracies between dark acoustic oscillation and intergalactic medium parameters}
\label{app:degeneracies}

\begin{table}[t]
\centering
%\begin{ruledtabular}
\begin{tabular}{lc}
Parameter & 95\% credible interval \\
\hline \hline 
$\kpeak\ [h/\mathrm{Mpc]}$ & $>5.104$ \\
$A$ & $< 1.376$ \\
$f$ & $< 0.243$ \\
%$f\,(0.5 < k_\mathrm{peak} < 1.5)$ & $< 0.268 $ \\
%$f\,(49.0 < k_\mathrm{peak} < 50.0)$ & $< 0.233$ \\
\hline \hline 
$\xi$ & unconstrained \\
$\tau_0(z=5.0)$ & $0.903 \quad 1.052$ \\
$\tau_0(z=4.6)$ & $0.933 \quad 1.095$ \\
$\tau_0(z=4.2)$ & $0.942 \quad 1.114$ \\
$n_\mathrm{s}$ & $0.957 \quad 0.971$ \\
$A_\mathrm{s}\times10^{9}$ & $2.045 \quad 2.144$ \\
$T_0(z=5.0)\ [{\rm K}]$ & $7074 \quad 14550$ \\
$T_0(z=4.6)\ [{\rm K}]$ & $7804 \quad 14383$ \\
$T_0(z=4.2)\ [{\rm K}]$ & $7615 \quad 14701$ \\
$\gamma(z=5.0)$ & $0.908 \quad 1.584$ \\
$\gamma(z=4.6)$ & $0.736 \quad 1.203$ \\
$\gamma(z=4.2)$ & $0.771 \quad 1.392$ \\
$u_0(z=5.0)\ [{\rm eV}\,m_p^{-1}]$ & $5.177 \quad 10.507$ \\
$u_0(z=4.6)\ [{\rm eV}\,m_p^{-1}]$ & $2.614 \quad 15.037$ \\
$u_0(z=4.2)\ [{\rm eV}\,m_p^{-1}]$ & $4.342 \quad 17.510$ \\
\end{tabular}
\caption{
1D marginalized 95\% credible intervals of the posterior distribution given Lyman-\(\alpha\) forest data.}
\label{tab:credible_intervals}
\end{table}

Figure \ref{figure:degen} shows 2D marginalized posterior distribution degeneracies between DAO transfer function and IGM parameters at \(z = 4.2\) (equivalent results are found at the other redshifts). 
The degeneracies are non-trivial since we have DAO parameters that both suppress and enhance the flux power spectrum in a scale-dependent way. As \(A\) increases, this increases small-scale flux power (see Fig.~\ref{figure:excluded}). This increase can be compensated by extra IGM heating from the UVB background that suppresses the small-scale power (making the IGM more diffuse), thus leading to the observed degeneracy between \(A\) and \(u_0\). We also find that hotter IGMs (higher \(u_0\)) are preferred when \(\kpeak\) decreases. This correlation occurs since the marginalized posterior is integrated over \(A\) (and all other parameters). The power boosting effect of \(A\) is stronger for lower \(\kpeak\) meaning that the increase in power is preferentially compensated by suppressing the power through extra IGM heating. This degeneracy diminishes as \(\kpeak\) increases, since the DAO peak in the linear transfer function (that causes the boost in flux power) gets pushed to higher \(k\) beyond the sensitivity of the data. Otherwise, the IGM posterior distributions are broadly consistent with the literature \citep[e.g.,][]{Rogers:2020ltq,Rogers_2022,Villasenor:2022aiy,Irsic:2023equ}, although we caution against a direct comparison to previous results given the extra dark matter parameters that we consider here which opens up new parameter degeneracies. The primordial power spectrum and effective optical depth posteriors are consistent with the prior distributions (see Table \ref{tab:credible_intervals}).

\section{Tests of patchy reionization and noise model}
\label{app:systematic_checks}

\begin{figure}
    \centering
    \includegraphics[trim={0cm 0cm 0cm 0cm},clip,width=\textwidth]{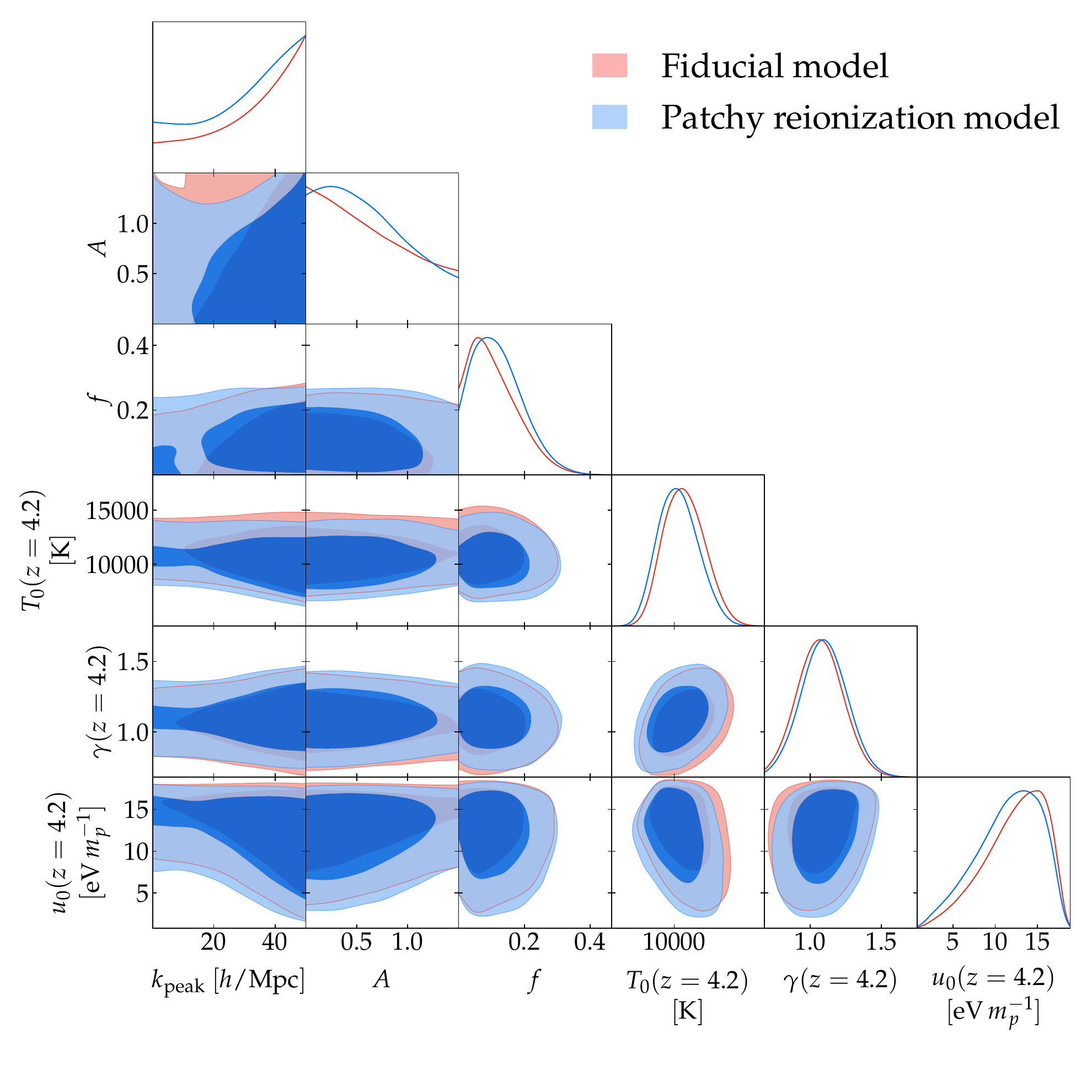}
    \caption{As Fig.~\ref{figure:constraints}, but showing degeneracies between DAO transfer function and IGM parameters at \(z=4.2\) for the fiducial analysis (red) and when accounting for a spatially-patchy reionization model \citep{Molaro_2021} (blue). Similar results are found at \(z=5\) and \(z=4.6\).}
    \label{figure:reion}
\end{figure}

In our simulations, we assume that reionization is a spatially-homogeneous process, i.e., the same (time-dependent) photoionization and photoheating rates are applied at each spatial position in the box. It is known that this is an approximation and that reionization proceeds by expanding ionizing bubbles around sources of ionizing sources, leading to large-scale (\(\sim 40\,\mathrm{Mpc}\)) spatial inhomogeneities during and for a time after reionization in the ionization and temperature fields \citep[e.g.,][]{Molaro_2021,Puchwein:2022wvk}. The scale of these inhomogeneities is much larger than the scale of the DAOs we test in this work. However, there is a secondary effect where the local mean flux will spatially fluctuate, leading to a \(\sim 10\%\) effect on the smallest scales we probe in the flux power spectrum as seen in radiative transfer simulations \citep{Chardin:2015uza,Wu:2019sgk}. This effect however is not significant compared to current data uncertainty. We explicitly test this effect by correcting the flux power spectra from the emulator according to the model presented in Ref.~\cite{Molaro_2021} that was calibrated to radiative transfer simulations. We find that, after applying this correction, there is no statistically significant change in the posterior distribution (see Fig.~\ref{figure:reion}) and that our results are robust to this effect given current statistical uncertainties. We anticipate, however, that the consideration of such effects will become increasingly important as the number of high-quality, high-resolution, high-redshift spectra increases, e.g., from the onset of extremely large telescopes \citep{Antypas:2022asj}.

\begin{figure}
    \centering
    \includegraphics[trim={0cm 0cm 0cm 0cm},clip,width=\textwidth]{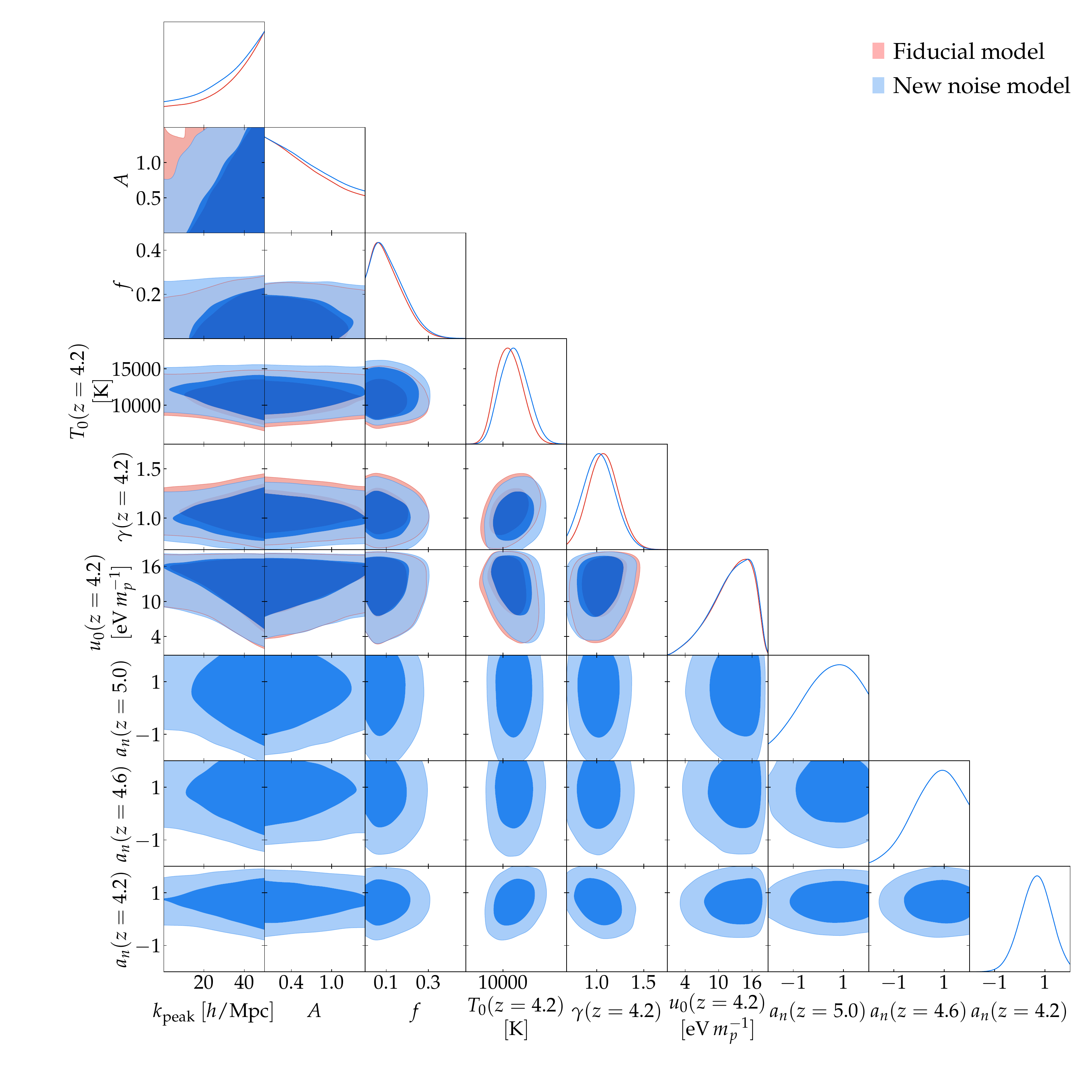}
    \caption{As Fig.~\ref{figure:constraints}, but comparing the fiducial analysis (red) and when allowing for a possible mis-modeling of the data noise \citep{Irsic:2023equ} (blue; introducing the additional parameters \(a_\mathrm{n}(z=z_i)\)).}
    \label{figure:noise_correction}
\end{figure}

As commented in the main letter, even the \(\Lambda\)CDM limit of the emulator does not return a good fit to the largest wavenumber bin of the data. This discrepancy was previously noted in Ref.~\cite{Irsic:2023equ}. They argue that this may arise from a mis-modeling of the noise in the data in the original data reduction \citep{Boera:2018vzq}, although this statement cannot be disentangled from any potential \textit{simulation} systematics. To account for this effect, they perform a test where they add a white noise term to the emulated flux power spectrum with a free amplitude parameter \(a_\mathrm{n}(z = z_i)\) that is varied. A different parameter for each redshift bin is allowed. They find that the posteriors on \(a_\mathrm{n}(z = z_i)\) mildly deviate from the no noise mis-modeling limit \(a_\mathrm{n}(z = z_i)=0\). We repeat this test in this analysis (see Fig.~\ref{figure:noise_correction}). While we find that \(a_\mathrm{n}(z = z_i)\) posteriors peak near unity, they remain consistent with the no mis-modeling limit also. We attribute the difference between the two noise analyses as to the effect of emulator uncertainty reducing the constraining power of the largest wavenumber data bin. Indeed, we do find that the fit to the data is improved with the addition of the \(a_\mathrm{n}\) parameters. Most importantly, we find that the DAO posteriors are very insensitive to the noise model. As discussed elsewhere, most of the constraining power on the DAO parameters comes from the larger affected   scales in these data, where the effect of the DAO amplitude \(A\) is strongest.
%%%%%%%%%%%%%%%%%%%%%%%
%%%%%%%%%%%%%%%%%%%%%%%
%%%%%%%%%%%%%%%%%%%%%%%
%%%%%%%%%%%%%%%%%%%%%%%
%%%%%%%%%%%%%%%%%%%%%%%

\end{document}